\documentclass[sigconf]{acmart}
\AtBeginDocument{%
  }
\renewcommand\footnotetextcopyrightpermission[1]{}
\acmISBN{978-1-4503-XXXX-X/2018/06}

\renewcommand\abstractname{\normalfont\textsc{Abstract}}

\newcommand{\calico}{\textsc{Calico}\xspace}

\usepackage{minted}
\usepackage{multirow}
\usepackage{adjustbox}
\usepackage{makecell}
\usepackage{booktabs}
\usepackage{lipsum} 
\usepackage{booktabs}
\usepackage{array}
\usepackage{fancyvrb}
\DefineVerbatimEnvironment{code}{Verbatim}{
  fontsize=\small,     
  xleftmargin=2em,     
  frame=leftline,        
  framesep=2pt,        
}
\usepackage{amsmath,amsopn}
\usepackage{endnotes,microtype,xspace,graphicx,fancyvrb}
\usepackage{booktabs}
\usepackage{makecell}
\usepackage{array,underscore,relsize}
\usepackage[english]{babel}
\usepackage{enumitem}
\usepackage{nicefrac}
\usepackage[labelfont=bf,font=small,skip=5pt]{caption}
\usepackage{subcaption}

\usepackage{fp}
\usepackage{siunitx}

\usepackage{balance}

\definecolor{linkcolor}{HTML}{647382} 
\definecolor{citecolor}{HTML}{647382}           
\definecolor{urlcolor}{rgb}{0.4,0.2,0.2}
\definecolor{sqlcolor}{HTML}{965d67}
\definecolor{smtcolor}{HTML}{5d968c}

\usepackage{afterpage}
\usepackage{graphicx, epsfig, endnotes}
\usepackage{colortbl}
\usepackage{tabularx}
\usepackage{ltablex}
\usepackage{listings}

\usepackage[ruled,vlined,linesnumbered]{algorithm2e}

\usepackage[capitalize,noabbrev,nameinlink]{cleveref}

\crefname{algocf}{alg.}{algs.}
\Crefname{algocf}{Algorithm}{Algorithms}

\usepackage{amsmath,amsopn}
\usepackage{stmaryrd}
\usepackage{mathtools}
\usepackage{mathrsfs}

\usepackage[scaled=0.8]{beramono}

\fvset{fontsize=\scriptsize,xleftmargin=8pt,numbers=left,numbersep=5pt}

\input{code/fmt}

\newcommand{\zerodisplayskips}{%
  \setlength{\abovedisplayskip}{0pt}%
  \setlength{\belowdisplayskip}{0pt}%
  \setlength{\abovedisplayshortskip}{0pt}%
  \setlength{\belowdisplayshortskip}{0pt}}
\appto{\normalsize}{\zerodisplayskips}
\appto{\small}{\zerodisplayskips}
\appto{\footnotesize}{\zerodisplayskips}

\def\Snospace~{\S{}}

\newif\ifdraft\drafttrue
\newif\ifnotes\notestrue
\ifdraft\else\notesfalse\fi

\input{glyphtounicode}
\newcolumntype{R}[1]{>{\raggedleft\let\newline\\\arraybackslash\hspace{0pt}}p{#1}}

\usepackage{siunitx}
\usepackage{tikz}

\newcommand*\circled[2][1.6]{\tikz[baseline=(char.base)]{
    \node[shape=circle, draw, inner sep=1pt, 
        minimum height={\f@size*#1},] (char) {\vphantom{WAH1g}#2};}}
\makeatother

\SetCommentSty{mycommfont}

\usepackage{xstring}
\newcommand{\PP}[1]{
\vspace{2px}
\noindent{\bf \IfEndWith{#1}{.}{#1}{#1.}}
}

\usepackage{xstring}
\newcommand{\PPP}[1]{
\vspace{2px}
\indent{\it \IfEndWith{#1}{.}{#1}{#1.}}
}

\newcommand{\boxbeg}{
\vspace{5px}
\noindent\begin{tabular}{|l|}\hline
\begin{minipage}{3.2in}
\vspace{5px}
\noindent
}

\newcommand{\boxend}{
\vspace{5px}
\end{minipage}\\ \hline
\end{tabular}
\vspace{1px}
}

\definecolor{light-gray}{gray}{0.95}

\makeatletter
\newcommand\BeraMonottfamily{%
  \def\fvm@Scale{0.85}
  \fontfamily{fvm}\selectfont
}
\makeatother
\lstdefinestyle{CStyle}{
language=C,
basicstyle=\BeraMonottfamily\footnotesize, 
keywordstyle=\color{sqlcolor}\bfseries,
literate = {-}{-}1, 
backgroundcolor = \color{light-gray},
xleftmargin=.25in,
xrightmargin=.25in,
morekeywords={ftruncate, open, mmap},
}
\lstdefinestyle{ScriStyle}{
    language=SQL,
    basicstyle=\BeraMonottfamily\footnotesize, 
    keywordstyle=\color{smtcolor}\bfseries,
    morekeywords={and, or, not},
    literate = {-}{-}1, 
}
\crefname{lstlisting}{listing}{listings}
\Crefname{lstlisting}{Listing}{Listings}

\definecolor{comment-color}{rgb}{0.25,0.25,0.25}

\definecolor{webgreen}{rgb}{0,.5,0}
\definecolor{webbrown}{rgb}{.6,0,0}
\definecolor{webblue}{rgb}{0,0,.7}

\newcommand{\squishitemize}{
 \begin{list}{$\bullet$}
  { \setlength{\itemsep}{0pt}
     \setlength{\parsep}{3pt}
     \setlength{\topsep}{3pt}
     \setlength{\partopsep}{0pt}
     \setlength{\leftmargin}{1.95em}
     \setlength{\labelwidth}{1.5em}
     \setlength{\labelsep}{0.5em} } }

\newcounter{Lcount}
\newcommand{\squishlist}{
    \begin{list}{\arabic{Lcount}. }
   { \usecounter{Lcount}
        \setlength{\itemsep}{0pt}
        \setlength{\parsep}{3pt}
        \setlength{\topsep}{3pt}
        \setlength{\partopsep}{0pt}
        \setlength{\leftmargin}{2em}
        \setlength{\labelwidth}{1.5em}
        \setlength{\labelsep}{0.5em} } }

\newcommand{\squishend}{\end{list}}

\newcommand{\captionTitle}[1]{\textbf{#1} --}

\definecolor{todo-color}{rgb}{1,0,0}

\newcommand{\parhead}[1]{%
  \par\smallskip\noindent\textbf{\textup{#1}}~\ignorespaces}

\renewcommand{\refname}{REFERENCES}

\begin{document}


\title{Making Array-Based Translation Practical for Modern, High-Performance Buffer Management}

%

\author{Xinjing Zhou}
\affiliation{%
  \institution{MIT CSAIL}
  \city{Cambridge}
  \state{Massachusetts}
  \country{USA}
  }
\email{xinjing@mit.edu}

\author{Jinming Hu}
\affiliation{
  \institution{sea-land.ai}  
  \city{}
  \country{China}
}

\email{conanhujinming@gmail.com}
\author{Andrew Pavlo}
\affiliation{
  \institution{Carnegie Mellon University}  
  \city{Pittsburgh}
  \state{Pennsylvania}
  \country{USA}
}
\email{pavlo@cs.cmu.edu}

\author{Michael Stonebraker}
\affiliation{
  \institution{MIT CSAIL} 
  \city{Cambridge}
  \state{Massachusetts}
  \country{USA}
}
\email{stonebraker@csail.mit.edu}




\setcounter{section}{0}
\begin{abstract}
Modern buffer pools must now support a broader workload mix than classic OLTP alone. In addition to
B-tree lookups, database systems increasingly serve scan-heavy analytics and vector-search indexes with 
irregular high-fan-out graph traversal access patterns. These workloads require a translation
mechanism---mapping logical page IDs to resident frames---that is simultaneously fast across these diverse access
patterns, deployable in user space,compatible with huge pages, 
easy to integrate 
, and still under DBMS control for eviction and I/O. Existing designs satisfy only subsets of these goals.

This paper presents \textbf{\calico}, a practical DBMS-controlled buffer pool built around array-based
translation, a decades-old-idea that was dissmissed but now viable with modern hardware. 



\calico decouples logical translation from OS page tables so that the DBMS can combine
low-overhead translation with huge-page-backed frames and fine-grained page management. To make
array translation practical and performant for DBMSes with large sparse hierarchical page identifiers,
\calico introduces three techniques: multi-level translation with path caching, hole punching for reclaiming cold translation
memory, and group prefetch to exploit parallelism.

Our evaluation across scans, OLTP-style B-tree accesses, and vector search shows that \calico matches or outperforms the existing state-of-the-art in-memory and 
out-of-memory performance. 
We also implement \calico as a drop-in replacement for
PostgreSQL's buffer manager and integrate it with \texttt{pgvector}. Across vector search,
and scan-heavy workloads, \calico delivers up to 3.9$\times$ in-memory and
6.5$\times$ larger-than-memory speedup for PostgreSQL vector search, speeds up scan-heavy queries
by up to 3$\times$. 

\end{abstract} 


\maketitle


\section{INTRODUCTION}
\label{sec:intro}
Disk-oriented database management systems (DBMSs) support data sets that exceed available physical
memory. A DBMS's buffer pool provides the key abstraction to achieve this
functionality.
It transparently handles the caching and eviction of pages in memory.
When the DBMS's internal components (e.g., query executors) access data
via logical page identifiers, they use the buffer pool to retrieve an in-memory buffer
frame for the requested page. Thus, the core operation of a buffer
pool is \textit{translation}: mapping a logical page ID to an in-memory frame. Translation sits on
the critical path of nearly every page access, so a buffer pool must keep it fast while
retaining fine-grained DBMS control over page I/O and eviction.

Although buffer pool design and implementation is an old topic in
databases~\cite{effelsberg1984principles}, there are recent proposals for high-performance buffer
pools~\cite{leis2018leanstore, umbra,vmcache,zhou2025practical, zinsmeister2026predictive}. These
newer designs primarily target OLTP-style workloads, which are dominated by B-tree traversals. The
rise of retrieval-augmented generation and semantic search is pushing DBMSs also to support vector
search~\cite{pgvector-hnsw,mysql}. Such workloads have access patterns that impose more stress on a
buffer pool than OLTP workloads~\cite{zhang2024there}: (1) partition-based indexes are
scan-heavy~\cite{jegou2010product,chen2021spann} and (2) graph-based indexes perform irregular,
high-fan-out traversals~\cite{malkov2018efficient,fu2017fast,jayaram2019diskann}. Hence, a modern
buffer pool must provide fast access for scan-heavy, B-tree, and graph workloads, and allow easy
integration with data structures to support future workload patterns.

The most salient design choice in a buffer pool implementation is whether the DBMS or OS manages the
page table(s)~\cite{crotty2022you}:


\paragraph{DBMS-managed:}
The most common approach is for the DBMS to maintain the buffer pool's
page table in user-space.
Most production DBMSs implement the buffer-page translation table as a user-space hash table that maps logical page identifiers to buffer frames~\cite{postgres,mysql,bridge1997oracle,sqlitepcache}. This design is popular because it preserves deployability and keeps eviction and I/O policy under DBMS control, but it introduces structural costs on the translation path: hash/probe work, pointer chasing, and synchronization, while hash functions also scatters adjacent page IDs. As a result, scan-heavy workloads lose prefetch efficiency, and high-parallelism graph workloads suffer from reduced memory-level parallelism. Predictive translation can improve hash-table translation~\cite{zinsmeister2026predictive}, but it is workload-dependent and still pays prediction plus lookup overhead. Pointer swizzling can remove explicit translation in some cases~\cite{graefe2014memory,leis2018leanstore,umbra}, yet it is invasive to data-structure internals and difficult to apply to graph-style pages with variable numbers of incoming references.

\paragraph{OS-managed:}
An alternative is for the DBMS to relinquish control of the memory to the OS.
OS page-table translation~\cite{vmcache,zhou2025practical} can make resident lookup
fast through hardware acceleration, but tie behavior to OS page-table management, weaken DBMS control, or require kernel changes
for good I/O performance, and interact poorly with huge pages under fine-grained page I/O and eviction.

Our systematic experimental analysis across scans, B-tree lookups, and graph traversal shows that no single existing mechanism is simultaneously strong for
scan-heavy, B-tree, and high-fan-out graph workloads while also supporting huge pages and
fine-grained DBMS I/O control. These limitations motivate a design goal that combines the strengths of both families: retain the full policy control of DBMS-managed translation state (eviction, recovery, and I/O scheduling) while approaching the translation efficiency of OS-managed page-table-based mechanisms.

Given this, we present \textbf{\calico}, a DBMS-managed buffer pool based on \textit{array
translation}: each logical page ID serves as an offset into a translation
array that maps to buffer frames. This design avoids hash probing and pointer chasing on the
translation path, keeps integration non-invasive via the existing PID-based interface, and decouples
logical translation from OS page tables. As a result, \calico can back frame memory with 2MB huge
pages for TLB efficiency while preserving fine-grained DBMS-managed eviction and I/O.

Array translation is not new. \citeauthor{effelsberg1984principles} discussed it in
1984~\cite{effelsberg1984principles}, but production systems have avoided it because large,
sparse, hierarchical page identifiers~\cite{pg-buffertag,mysql-pageid,sqlserver-page-id,
rocksdb-block-cache-key-format,oracle-page-id} made flat arrays too expensive in terms of memory
overhead. \calico makes the approach practical and performant through three techniques:
\textbf{multi-level translation} with path caching for managing sparse hierarchical IDs
(\cref{sec:multi-level-calico}), \textbf{active hole punching} to reclaim cold translation memory
(\cref{sec:lazy-allocation}), and \textbf{group prefetch} to hide indirection latency during
high-fan-out graph traversal (\cref{sec:implementation}). On modern hardware, these mechanisms let
array translation match OS-based page table translation performance while substantially
outperforming hash-table translation across scans, B-tree lookups, and graph traversal.

We implement \calico as a drop-in replacement for PostgreSQL's buffer manager with about 2.6K lines
of code changes. \calico improves \texttt{pgvector} by 2.9--3.95$\times$,
speeds up scan-heavy PostgreSQL queries by up to 3$\times$, and outperforms the
in-memory-only libraries such as Faiss~\cite{Faiss} when data is resident.

Our contributions are:
\begin{itemize}[leftmargin=*]
    \item \textbf{A workload-grounded analysis of buffer-pool translation.}
    We identify the access regimes that now matter for modern buffer pools---scan-heavy access,
    B-tree lookups, and high-fan-out graph traversal---and provide a systematic comparison of
    hash-table translation, predictive translation, pointer swizzling, OS page-table translation,
    and array translation across these regimes (\cref{sec:mot,sec:translation}).

    \item \textbf{A practical DBMS-managed array-translation buffer pool.}
    We present \calico, which decouples logical translation from OS page tables so that the DBMS can
    combine low-overhead translation, huge-page-backed frames, and fine-grained eviction/I/O
    control in one design (\cref{sec:design}).

    \item \textbf{Techniques that make array translation practical and effective.}
    \calico introduces multi-level translation for sparse hierarchical page IDs, active hole
    punching to reclaim cold translation regions, and group prefetch plus optimistic reads to
    exploit memory-level parallelism in graph traversal and vector search
    (\cref{sec:design,sec:implementation}).

    \item \textbf{An implementation in PostgreSQL and a broad end-to-end evaluation.}
    We implement \calico as a drop-in replacement for PostgreSQL's buffer manager, integrate it
    with \texttt{pgvector}, and evaluate it on vector search, OLTP, scan-heavy PostgreSQL queries,
    translation-memory overhead, ablation, and cross-platform experiments. \calico delivers up to
    3.95$\times$ in-memory and 6.57$\times$ larger-than-memory speedup for PostgreSQL vector search
    while also improving scan-heavy and OLTP workloads (\cref{sec:eval,sec:eval-pg}).
\end{itemize}

\begin{table*}[t]
    \centering
    \caption{
        \captionTitle{Key Properties and Workload Efficiency of Buffer-Pool Designs}
        Comparison of the three translation-data-structure classes. Workload-efficiency cells report
        SS, RS, PL, and GT using color-coded badges, where green, yellow, and red indicate strong,
        mixed, and weak efficiency. A \textbackslash\ indicates not applicable because pointer
        swizzling and predictive translation require DBMS control of the translation structure.
        GT$^{\dagger}$ in the pointer-swizzling column denotes limited support for graph-style
        workloads due to multi-parent references. The
        property columns summarize huge-page friendliness, required OS modifications, I/O control,
        and memory-overhead scaling.
    }
    \label{tab:buffer-comparison}
    {\small \newcommand{\effstrong}[1]{\begingroup\setlength{\fboxsep}{1pt}\colorbox{green!20}{\textsf{\scriptsize #1}}\endgroup}
\newcommand{\effmedium}[1]{\begingroup\setlength{\fboxsep}{1pt}\colorbox{yellow!30}{\textsf{\scriptsize #1}}\endgroup}
\newcommand{\effweak}[1]{\begingroup\setlength{\fboxsep}{1pt}\colorbox{red!20}{\textsf{\scriptsize #1}}\endgroup}
\newcommand{\effrow}[4]{#1\,#2\,#3\,#4}
\newcommand{\propstrong}[1]{\begingroup\setlength{\fboxsep}{1pt}\colorbox{green!20}{\textsf{\scriptsize #1}}\endgroup}
\newcommand{\propmixed}[1]{\begingroup\setlength{\fboxsep}{1pt}\colorbox{yellow!30}{\textsf{\scriptsize #1}}\endgroup}
\newcommand{\propweak}[1]{\begingroup\setlength{\fboxsep}{1pt}\colorbox{red!20}{\textsf{\scriptsize #1}}\endgroup}
\setlength{\tabcolsep}{4pt}
\renewcommand{\arraystretch}{1.12}
\begin{tabular*}{\textwidth}{@{\extracolsep{\fill}}llccccccc@{}}
\toprule
\multirow{2}{*}{\makecell[l]{}}
& \multirow{2}{*}{\makecell[l]{\textbf{Where}\\\textbf{Stored}}}
& \multicolumn{4}{c}{\textbf{Key Properties}}
& \multicolumn{3}{c}{\textbf{Workload Efficiency}} \\
\cmidrule(lr){3-6}\cmidrule(l){7-9}
& & \makecell[c]{\textbf{Huge-page}\\\textbf{friendly}}
& \makecell[c]{\textbf{OS mods}}
& \makecell[c]{\textbf{I/O control}}
& \makecell[c]{\textbf{Memory}\\\textbf{overhead}}
& \makecell[c]{\textbf{Baseline}}
& \makecell[c]{\textbf{+ Predictive}\\\textbf{Translation}}
& \makecell[c]{\textbf{+ Pointer Swizzling}}
\\
\midrule
\makecell[r]{\textbf{Hash Table Translation}}
& \makecell[l]{Userspace}
& \propstrong{yes}
& \propstrong{none}
& \propstrong{full}
& \makecell[c]{\effstrong{\textit{O}(\# cached pages)}}
& \makecell[c]{\effrow{\effweak{SS}}{\effmedium{RS}}{\effmedium{PL}}{\effweak{GT}}}
& \makecell[c]{\effrow{\effmedium{SS}}{\effstrong{RS}}{\effstrong{PL}}{\effmedium{GT}}}
& \makecell[c]{\effrow{\effstrong{SS}}{\effstrong{RS}}{\effstrong{PL}}{\effweak{GT$\dagger$}}}
\\
\addlinespace[2pt]

\makecell[r]{\textbf{Hardware-accelerated Radix Tree}}
& \makecell[l]{OS Kernel}
& \propweak{no}
& \propmixed{mixed}
& \propmixed{mixed}
& \makecell[c]{\effmedium{\textit{O}(\# storage pages)}}
& \makecell[c]{\effrow{\effstrong{SS}}{\effstrong{RS}}{\effstrong{PL}}{\effmedium{GT}}}
& \makecell[c]{\textbackslash}
& \makecell[c]{\textbackslash}
\\
\addlinespace[2pt]

\makecell[r]{\textbf{Array Translation (\calico)}}
& \makecell[l]{Userspace}
& \propstrong{yes}
& \propstrong{none}
& \propstrong{full}
& \makecell[c]{\effmedium{\textit{O}(\# storage pages)}}
& \makecell[c]{\effrow{\effstrong{SS}}{\effstrong{RS}}{\effstrong{PL}}{\effstrong{GT}}}
& \makecell[c]{\effrow{\effstrong{SS}}{\effstrong{RS}}{\effstrong{PL}}{\effstrong{GT}}}
& \makecell[c]{\effrow{\effstrong{SS}}{\effstrong{RS}}{\effstrong{PL}}{\effweak{GT$\dagger$}}}
\\

\specialrule{\heavyrulewidth}{0.25ex}{0ex}
\end{tabular*}
}
\end{table*}

\section{BACKGROUND AND MOTIVATION}
\label{sec:mot}
We now characterize the workload requirements that matter in modern workloads and use them to
motivate a design-space decomposition of buffer-pool translation.

\subsection{Workload and Operational Requirements}

\parhead{Sequential scan (SS)} streams through long contiguous PID ranges (e.g., heap scans,
partitioned-based/brute-force vector search~\cite{jegou2010product,chen2021spann}). SS tests whether translation preserves
spatial locality and lets hardware prefetchers run ahead.

\parhead{Range scan (RS)} starts with a small dependent lookup (e.g., B-tree scan) and
then scans consecutive leaf or posting-list pages. It combines short pointer-dependent phases with
long near-sequential translation phases.

\parhead{Point lookup (PL)} captures latency-sensitive random access such as a B-tree root-to-leaf
descent or key-value probe. Here, per-access translation latency is important.

\parhead{Graph traversal (GT)} captures irregular, high-memory-level-parallelism traversals such as HNSW search~\cite{malkov2018efficient,fu2017fast,jayaram2019diskann}. One
visited node may expose many candidate neighbors at once, so translation must preserve parallelism.

These four patterns span the locality and parallelism regimes that modern buffer pools must handle.
We use SS, RS, PL, and GT throughout the paper and use them to explain the trade-offs of each base
design family in the next subsection.

\subsection{Design Space of Buffer Pools}

Buffer pool designs differ along two orthogonal axes: the \textbf{translation data structure}, which
determines how logical page IDs map to resident frames, and \textbf{where translation state is
stored}. The first axis distinguishes \emph{Hash Table Translation},\emph{Hardware Page Table Translation}, and \emph{Array Translation}. The second axis distinguishes \emph{Userspace page tables} and
\emph{OS page tables}. This separation makes the base
design choice explicit before discussing optional optimizations.

\cref{tab:buffer-comparison} is organized as follows: rows are the \emph{base translation mechanisms}.
Key-property columns summarize system properties, and workload columns report baseline behavior plus
two orthogonal techniques (\emph{+ Predictive Translation}, \emph{+ Pointer Swizzling}).

\parhead{DBMS-managed hash-table pools:} Hash table translation is the most widely used base design in production
systems~\cite{narasayya2015sharing,postgres,mysql,sqlitepcache,wiredtiger}. A hash table maps page
identifiers (PIDs) to buffer frames, giving the DBMS full control over eviction policies and I/O
scheduling. The trade-off is critical-path overhead: collision resolution, synchronization, and pointer
chasing create dependency chains that can limit memory-level parallelism for high-fanout workloads.
Hashing also inherently scatters consecutive PIDs, destroying spatial locality on scans; under high thread counts,
lock/atomic synchronization increases coherence traffic.

\parhead{OS-managed page-table translation: } This approach stores translation state in OS/MMU page
tables~\cite{vmcache,zhou2025practical}. \textit{File-backed \texttt{mmap}} delegates page faults,
caching, and eviction to the OS~\cite{henry2019howard,ravendb,sqlitepcache}, reducing DBMS control
over replacement and I/O scheduling. \textit{Virtual-memory-based} systems such as vmcache keep
translation in page tables but move policy into the DBMS. In \textbf{vmcache} (without exmap), the
DBMS performs explicit I/O and eviction
using standard kernel interfaces, which preserves user-space deployability but pays higher per-page
VM-management overhead such as TLB-shootdown, especially on high-end storage devices. 
Kernel modifications are required to improve scalability of page fault handling and eviction, 
as in \textbf{vmcache+exmap}~\cite{vmcache} and \textbf{Tabby+libdbos}~\cite{zhou2025practical}, 
but such modifications face deployment barriers.

\parhead{Eviction granularity mismatch:}
OS-managed page-table translation fundamentally shares a limitation.
DBMSs need fine-grained page management and I/O (typically 4--32KB), while huge pages (2MB) are
important for TLB efficiency. When translation state resides in hardware page tables, the OS cannot
cleanly evict a single 4KB subpage from a mapped 2MB huge page; it must evict the full 2MB
region~\cite{hugepageswap}, split the huge page~\cite{michailidis2019mega,manocha2023architectural},
or fail/refuse the operation~\cite{madvisemanual,madvisedontneed}. Hence these systems must choose
between 4KB pages (fine-grained eviction and I/O, higher TLB pressure) and 2MB pages (lower TLB
pressure, high I/O amplification).

\parhead{Pointer swizzling and predictive translation:} Conceptually, they can
be layered on top of any mechanism that exposes a modifiable translation path, but in practice they are
most applicable to DBMS-managed translation structures; for OS-managed page-table translation, the
structure is fixed in hardware/OS page tables. Pointer swizzling~\cite{white1994pointer,graefe2014memory,leis2018leanstore,umbra,leis2024leanstore}
removes lookup overhead by replacing logical IDs with direct pointers, but introduces invasive coupling:
eviction requires unswizzling/validation bookkeeping, which is hard for multi-parent references (e.g.,
graph fan-in, sibling links, secondary-index backlinks). LIPAH~\cite{LIPAH} improves coverage but still
requires in-page hints and validation logic. Predictive translation~\cite{zinsmeister2026predictive}
keeps a hash-table translation layer but assigns pages preferred frame positions so the CPU can
speculatively access the likely frame while translation is still in progress. 
It is effective when hot pages remain in preferred positions, but benefits are workload-dependent and can shrink under irregular access
or frequent mispredictions.

\parhead{Array Translation:} \calico pursues this underexplored DBMS-managed design
point in \cref{tab:buffer-comparison}. It retains DBMS control and non-invasive integration, but
replaces associative lookup with direct array translation to improve locality and translation efficiency. Despite being discussed
historically~\cite{effelsberg1984principles}, this point has seen limited deployment in modern buffer
pools. The next section therefore performs a focused translation-performance analysis to establish
whether array translation can match the fast path of hardware-assisted alternatives while preserving
DBMS-managed control.

\section{TRANSLATION PERFORMANCE ANALYSIS}
\label{sec:translation}

\begin{figure*}[t]
    \centering
        \includegraphics[width=1\textwidth]{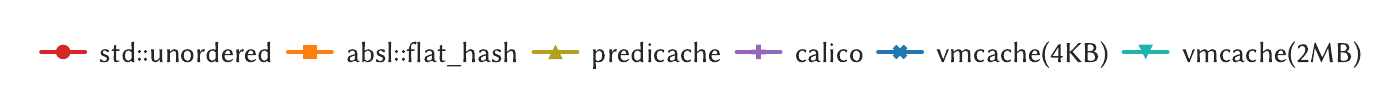}
     \\
     \begin{subfigure}{0.24\textwidth}
        \centering
        \includegraphics[width=\textwidth]{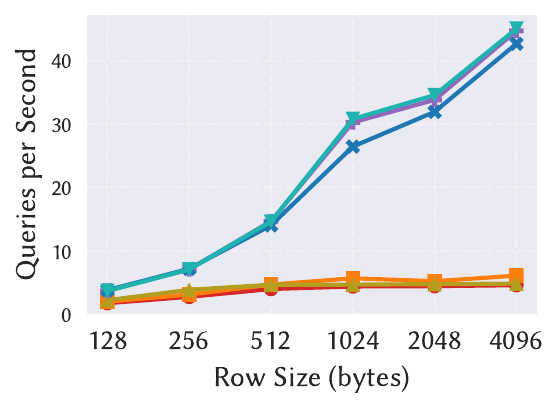}
        \caption{Sequential scan (heap scan)}
        \label{fig:olap-sequential}
     \end{subfigure}
     \hfill
     \begin{subfigure}{0.24\textwidth}
        \centering
        \includegraphics[width=\textwidth]{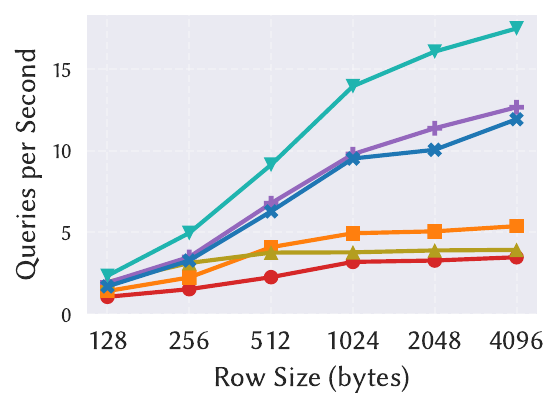}
        \caption{Random range scan (B-tree scan)}
        \label{fig:olap-random}
     \end{subfigure}
     \hfill
     \begin{subfigure}{0.24\textwidth}
        \centering
        \includegraphics[width=\textwidth]{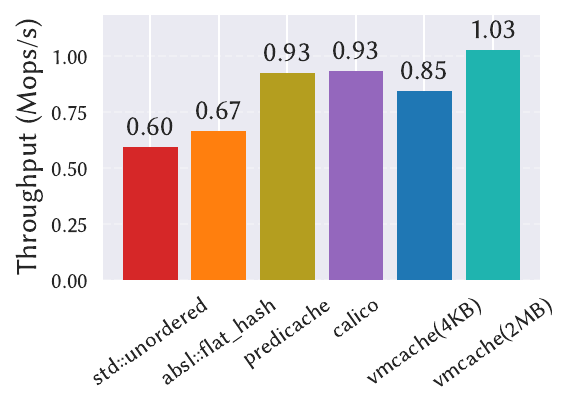}
        \caption{B-tree random point lookup}
        \label{fig:btree-single}
     \end{subfigure}
     \hfill
     \begin{subfigure}{0.24\textwidth}
        \centering
        \includegraphics[width=\textwidth]{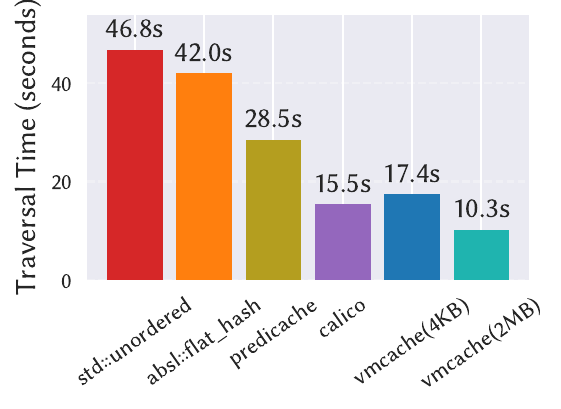}
        \caption{Graph BFS traversal}
        \label{fig:graph-bfs-1t}
     \end{subfigure}
    \Description{Four plots compare translation overhead across workloads. Sequential scan and random scan show array-based translation outperforming hash tables, with predicache helping little on scans. B-tree lookup shows predicache close to array on predictable tree traversals. Graph BFS shows array fastest, vmcache next, predicache intermediate, and hash tables slowest.}
    \caption{
        \captionTitle{Translation Overhead Across Workloads}
        (a) Sequential scan: Hash tables destroy spatial locality, causing 6.9$\times$ slowdown vs
        array/vmcache; predicache provides no benefit for sequential access.
        (b) Random range scan: Gap persists at 2.0$\times$ despite random access; predicache
        underperforms plain hash tables.
        (c) B-tree lookup: Hash tables cause 1.56$\times$ slowdown; predicache matches array
        by overlapping hash table access with buffer frame access through CPU speculative execution.
        (d) Graph BFS: Translation serialization limits memory-level parallelism, causing 
3.4$\times$
        slowdown; predicache helps partially (1.5$\times$) but irregular traversal limits
        speculation accuracy. Array-based translation matches or outperforms vmcache with 4KB
        pages while retaining fine-grained I/O.
    }
    \label{fig:translation-overhead}
\end{figure*}

To quantify the overhead of different translation mechanisms, we compare the four base
approaches from \cref{sec:mot}: hash tables, PrediCache~\cite{zinsmeister2026predictive},
vmcache~\cite{vmcache} (4KB and 2MB pages), and calico's array-based translation. All implement the same
interface: given an 8-byte page ID, return a pointer to the buffer frame.

\parhead{Setup.} We run on a m7a.8xlarge AWS EC2 instance (AMD EPYC 9R14). All experiments are
single-threaded with data fully resident, isolating translation overhead from eviction and
synchronization. Page IDs are allocated from 0 sequentially; hash table, PrediCache, and array
configurations use huge pages for buffer frames.

We evaluate three representative workloads that stress different aspects of translation performance:
\textbf{(1) Scan workloads} (\cref{sec:scan-workloads}) measure how well each mechanism preserves spatial locality for sequential and random page access patterns common in analytical queries and partition-based vector search.
\textbf{(2) B-tree lookup} (\cref{sec:btree-lookup}) evaluates translation overhead for the canonical OLTP workload with dependent memory accesses that limit parallelism.
\textbf{(3) Graph traversal} (\cref{sec:graph-bfs}) stresses memory-level parallelism with highly parallel random accesses representative of proximity-graph-based vector search.

\subsection{Scan-heavy access}
\label{sec:scan-workloads}

\begin{table}[t]
    \centering
    \caption{
        \captionTitle{Microarchitecture Metrics}
        Performance counters per page scanned (8KB page, 1024-byte rows) for sum
        query on 50GB buffer pool scanning 5GB data. Sequential scan: consecutive page IDs (e.g.,
        heap scan); Random scan: non-consecutive page IDs (e.g., B-tree leaf scan). Metrics
        averaged across all pages scanned.
    }
    \label{tab:olap-uarch}
    {\footnotesize \begin{tabular}{l@{\hskip 0.08in}r@{\hskip 0.05in}r@{\hskip 0.05in}r@{\hskip 0.05in}r@{\hskip 0.05in}r@{\hskip 0.05in}r@{\hskip 0.05in}r}
\toprule
\textbf{Config} & \textbf{QPS} & \textbf{IPC} & \textbf{Inst} & \textbf{LLC Ref} & \textbf{LLC Miss} & \textbf{DTLB Miss} & \textbf{Cycles} \\
\midrule
\multicolumn{8}{l}{\textit{Sequential Scan (heap scan)}} \\
\texttt{std::unordered} & 4.38 & 0.15 & 115 & 21.2 & 3.09 & 1.11 & 755 \\
\texttt{absl::flat\_hash} & 5.67 & 0.21 & 122 & 18.9 & 2.53 & 0.81 & 585 \\
\texttt{predicache} & 4.67 & 0.17 & 118 & 23.5 & 7.55 & 0.54 & 702 \\
\texttt{calico} & 30.3 & 0.77 & 83.8 & 13.8 & 1.33 & 0.002 & 109 \\
\texttt{vmcache(4KB)} & 26.4 & 0.64 & 80.1 & 14.2 & 1.75 & 1.25 & 125 \\
\texttt{vmcache(2MB)} & 30.8 & 0.74 & 80.1 & 14.0 & 1.42 & 0.002 & 108 \\
\midrule
\multicolumn{8}{l}{\textit{Random Scan (B-tree leaf scan)}} \\
\texttt{std::unordered} & 3.20 & 0.15 & 155 & 34.0 & 10.9 & 2.35 & 1033 \\
\texttt{absl::flat\_hash} & 4.95 & 0.24 & 161 & 29.1 & 8.09 & 1.58 & 669 \\
\texttt{predicache} & 3.78 & 0.14 & 119 & 27.4 & 9.43 & 1.01 & 870 \\
\texttt{calico} & 9.78 & 0.36 & 123 & 22.5 & 5.22 & 0.56 & 338 \\
\texttt{vmcache(4KB)} & 9.52 & 0.34 & 120 & 20.8 & 7.36 & 2.28 & 347 \\
\texttt{vmcache(2MB)} & 13.9 & 0.50 & 120 & 20.7 & 5.01 & 0.56 & 237 \\
\bottomrule
\end{tabular}
}
\end{table}

We measure spatial locality preservation using a scan query on heap pages with row-oriented
format (\cref{fig:olap-sequential}, \cref{fig:olap-random}, \cref{tab:olap-uarch}). Sequential scans
are common in analytical queries and table scans, while random scans occur during index-organized
table access. We implement an aggregation query summing over a column and scanning 5GB of 8KB pages
(1024-byte rows). \textit{Sequential} scans access consecutive page IDs (simulating heap scans),
while \textit{random} scans access non-consecutive page IDs (simulating B-tree leaf scans). This
workload isolates the overhead that translation mechanisms introduce for scan-heavy workloads.

As discussed in \cref{sec:mot}, hash tables scatter consecutive PIDs, destroying spatial locality.
Array indexing preserves adjacency: consecutive PIDs map to consecutive 8-byte entries, enabling
hardware prefetching. Results show \texttt{array} achieves 30.3 QPS for sequential scans,
6.9$\times$ faster than \texttt{std::unordered} and 6.5$\times$ faster than \texttt{predicache},
while essentially matching \texttt{vmcache(2MB)} at 30.8 QPS (\cref{tab:olap-uarch}). It also cuts
LLC misses to 1.33 per page, versus 2.53 for \texttt{absl::flat\_hash} and 7.55 for
\texttt{predicache}. For random scans, \texttt{array} reaches 9.78 QPS, still 2.0$\times$ faster
than \texttt{absl::flat\_hash}, 2.6$\times$ faster than \texttt{predicache}, and slightly ahead of
\texttt{vmcache(4KB)} at 9.52 QPS. Overall, array translation stays close to vmcache with huge
pages while retaining fine-grained eviction.

A key reason is \textbf{cache efficiency}: array entries store only the 8-byte frame ID (the page
ID is implicit in the index), whereas hash tables must store both keys and values (16+ bytes per
entry) and maintain load factors of 0.5--0.9, wasting space to control collisions. This density
allows more translation entries per cache line, which the microarchitecture data confirms:
\texttt{array} incurs only 1.33/4.64 LLC misses per page for sequential/random scans, versus
2.53/5.28 for \texttt{absl::flat\_hash}.

PrediCache reduces instruction count relative to conventional hash tables, but its scan performance
remains bounded by the cache-unfriendly layout of the underlying hash table. It
still incurs 23.5/27.4 LLC references and 7.55/9.43 LLC misses per page for sequential/random
access, leading to 702/870 cycles per page and leaving it below even \texttt{absl::flat\_hash} on
both scans. Speculative execution can overlap part of a lookup, but it cannot restore the spatial
locality that hashing destroys across a scan stream.
\subsection{Point lookup}
\label{sec:btree-lookup}

\begin{table}[t]
\centering
\caption{\textbf{Microarchitecture metrics per B-tree lookup for YCSB-C workload on 24GB buffer pool with 100M records (128 bytes each). Single-threaded read-only point queries with typical tree depth 4--5 nodes. Metrics averaged across all lookups.}}
\label{tab:btree-uarch}
\footnotesize
\begin{tabular}{l@{\hskip 0.08in}r@{\hskip 0.05in}r@{\hskip 0.05in}r@{\hskip 0.05in}r@{\hskip 0.05in}r@{\hskip 0.05in}r@{\hskip 0.05in}r}
\toprule
\textbf{Config} & \textbf{Mops/s} & \textbf{IPC} & \textbf{Inst} & \textbf{LLC Ref} & \textbf{LLC Miss} & \textbf{DTLB Miss} & \textbf{Cycles} \\
\midrule
\texttt{std::unordered} & 0.60 & 0.32 & 1168 & 25.4 & 4.10 & --- & 3,615 \\
\texttt{absl::flat\_hash} & 0.67 & 0.37 & 1216 & 21.2 & 3.35 & --- & 3,246 \\
\texttt{predicache} & 0.93 & 0.55 & 1273 & 15.9 & 1.30 & --- & 2,332 \\
\texttt{calico} & 0.93 & 0.49 & 1142 & 12.5 & 0.46 & --- & 2,315 \\
\texttt{vmcache(4KB)} & 0.85 & 0.40 & 1009 & 16.8 & 4.40 & --- & 2,536 \\
\texttt{vmcache(2MB)} & 1.03 & 0.48 & 1010 & 14.8 & 1.08 & --- & 2,097 \\
\bottomrule
\end{tabular}
\end{table}

B-tree point queries represent the core OLTP workload. Each lookup traverses from root to leaf, 
accessing random pages with dependent loads that limit memory-level parallelism. We use YCSB-C 
(read-only) on 100M records (128 bytes each, 24GB buffer pool, 4KB pages) to evaluate translation 
overhead in this common access pattern with results in \cref{fig:btree-single} and 
\cref{tab:btree-uarch}.

Chained hash tables (\texttt{std::unordered}) suffer from pointer chasing. The dependent memory 
operations serialize execution, causing the CPU to spend most cycles stalled. It achieves only 0.60 
Mops/s with high LLC misses (25.4/op). Open-addressing fast hash table \texttt{absl::flat\_hash} 
improves cache efficiency and reaches 0.67 Mops/s. \texttt{predicache} reaches 0.93 Mops/s because
B-tree's predictable root-to-leaf path enables accurate speculation that hides hash table latency.
\texttt{Array} matches predicache at 0.93 Mops/s with fewer instructions (1142 vs 1273) and lower
LLC misses (12.5 vs 15.9), achieving the same throughput through simpler translation. Note that
array retains fine-grained eviction control that is impossible with huge pages in vmcache.

\subsection{Graph traversal}
\label{sec:graph-bfs}

\begin{table}[t]
\centering
\caption{\textbf{Microarchitecture metrics per graph node visited (including neighbor probes) for BFS traversal on 5M-node proximity graph. Single-threaded traversal from same starting node. Metrics averaged across all graph nodes visited during full graph traversal.}}
\label{tab:graph-uarch}
\footnotesize
\begin{tabular}{l@{\hskip 0.08in}r@{\hskip 0.05in}r@{\hskip 0.05in}r@{\hskip 0.05in}r@{\hskip 0.05in}r@{\hskip 0.05in}r@{\hskip 0.05in}r}
\toprule
\textbf{Config} & \textbf{Time (s)} & \textbf{IPC} & \textbf{Inst} & \textbf{LLC Ref} & \textbf{LLC Miss} & \textbf{DTLB Miss} & \textbf{Cycles} \\
\midrule
\texttt{std::unordered} & 46.8 & 0.18 & 3552 & 548 & 337 & 110 & 20,206 \\
\texttt{absl::flat\_hash} & 42.0 & 0.22 & 3937 & 502 & 255 & 91 & 18,107 \\
\texttt{predicache} & 28.5 & 0.36 & 4385 & 307 & 183 & 51 & 12,335 \\
\texttt{calico} & 15.5 & 0.63 & 2248 & 265 & 142 & 60 & 6,671 \\
\texttt{vmcache(4KB)} & 17.4 & 0.28 & 2082 & 261 & 147 & 39 & 7,476 \\
\texttt{vmcache(2MB)} & 10.3 & 0.48 & 2079 & 175 & 114 & 27 & 4,367 \\
\bottomrule
\end{tabular}
\end{table}

Graph BFS traversal exposes translation's impact on memory-level parallelism, which is an
important property for modern vector search workloads. We traverse a 5M-node graph (simulating HNSW
vector index beam search) where each node connects to 44 random neighbors (4KB page per node). When
visiting a node, the algorithm probes all neighbors. The workload is deliberately compute-light to
maximize buffer pool stress and expose translation bottlenecks. The results are shown in
\cref{fig:graph-bfs-1t} and \cref{tab:graph-uarch}.

For chained hash tables, each neighbor access requires dependent pointer chasing through
buckets, preventing the CPU from issuing parallel loads. Even advanced open-addressing tables suffer
because SIMD probing does not help much. \texttt{std::unordered} takes 46.8s with very low IPC
(0.18) and high cache/TLB misses (337 LLC misses, 110 DTLB misses per node), while
\texttt{absl::flat\_hash} improves only slightly to 42.0s.

\texttt{predicache} completes in 28.5s (1.5$\times$ faster than \texttt{absl::flat\_hash}),
but unlike B-tree's predictable root-to-leaf path, graph traversal order depends on distance
computations over all neighbors, limiting speculation accuracy and leaving \texttt{predicache}
still 1.8$\times$ slower than \texttt{array}.
\texttt{Array} completes in 15.5s (2.7$\times$ faster than \texttt{absl::flat\_hash}), outperforming even
\texttt{vmcache} with 4KB pages. Direct array indexing lets the CPU issue all neighbor translations
in parallel without data dependencies, fully exploiting memory-level parallelism.

\subsection{Takeaway}
\label{sec:array-takeaway}

Across all four workloads, array-based translation matches or outperforms vmcache with 4KB pages. Its advantage over hash tables stems from two
reinforcing effects: (1) eliminating hash computation, pointer chasing, and synchronization removes
data dependencies that serialize memory accesses, and (2) higher cache density keeps more of the translation state working
set in the CPU cache hierarchy. These effects are most pronounced for high-MLP graph traversal
(2.7$\times$) and sequential scans (6.9$\times$), but also measurable for point lookups
(1.4$\times$). Arrays are therefore fast enough to be a serious design point.


\begin{figure}[t]
    \begin{subfigure}{0.6\textwidth}
    \includegraphics[scale=0.35]{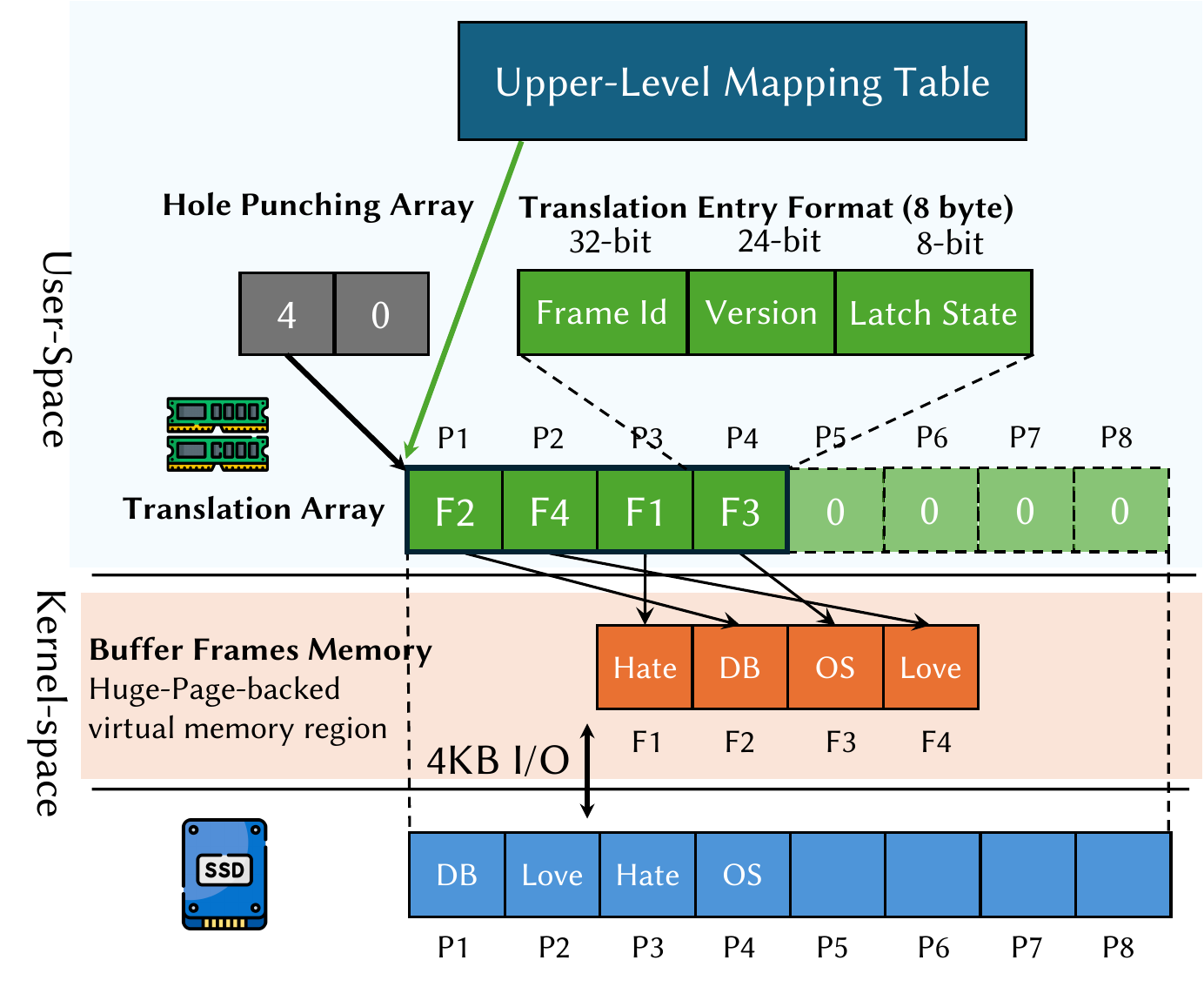}
    \end{subfigure}
    \Description{Architecture diagram of \calico showing an upper-level mapping table, DBMS-managed logical translation state in last-level translation arrays, OS-managed huge-page-backed frame memory, and a hole-punching reference-count array. For readability, each metadata group in the figure covers four translation entries (32 bytes); in practice, groups are OS-page sized (typically 4KB of translation entries).}
    \caption{\captionTitle{\calico Buffer Manager Architecture}
    \calico separates DBMS-managed logical translation control from OS-managed physical backing.
    The upper-level mapping table resolves page-ID prefixes to last-level translation arrays. Each
    64-bit translation entry encodes frame ID, version, and latch state. Frame memory is huge-page-backed
    for TLB efficiency. The hole-punching reference-count array tracks groups of translation entries so
    cold regions of the translation array can be reclaimed without affecting frame-memory mappings.
    For readability,
    the figure shows 4-entry groups (32 bytes); in practice, groups are typically one OS page of
    translation entries (4KB).
    }
    \label{fig:CalicoArch}
\end{figure}

\begin{figure}[t]
    \centering
    \includegraphics[width=0.5\textwidth]{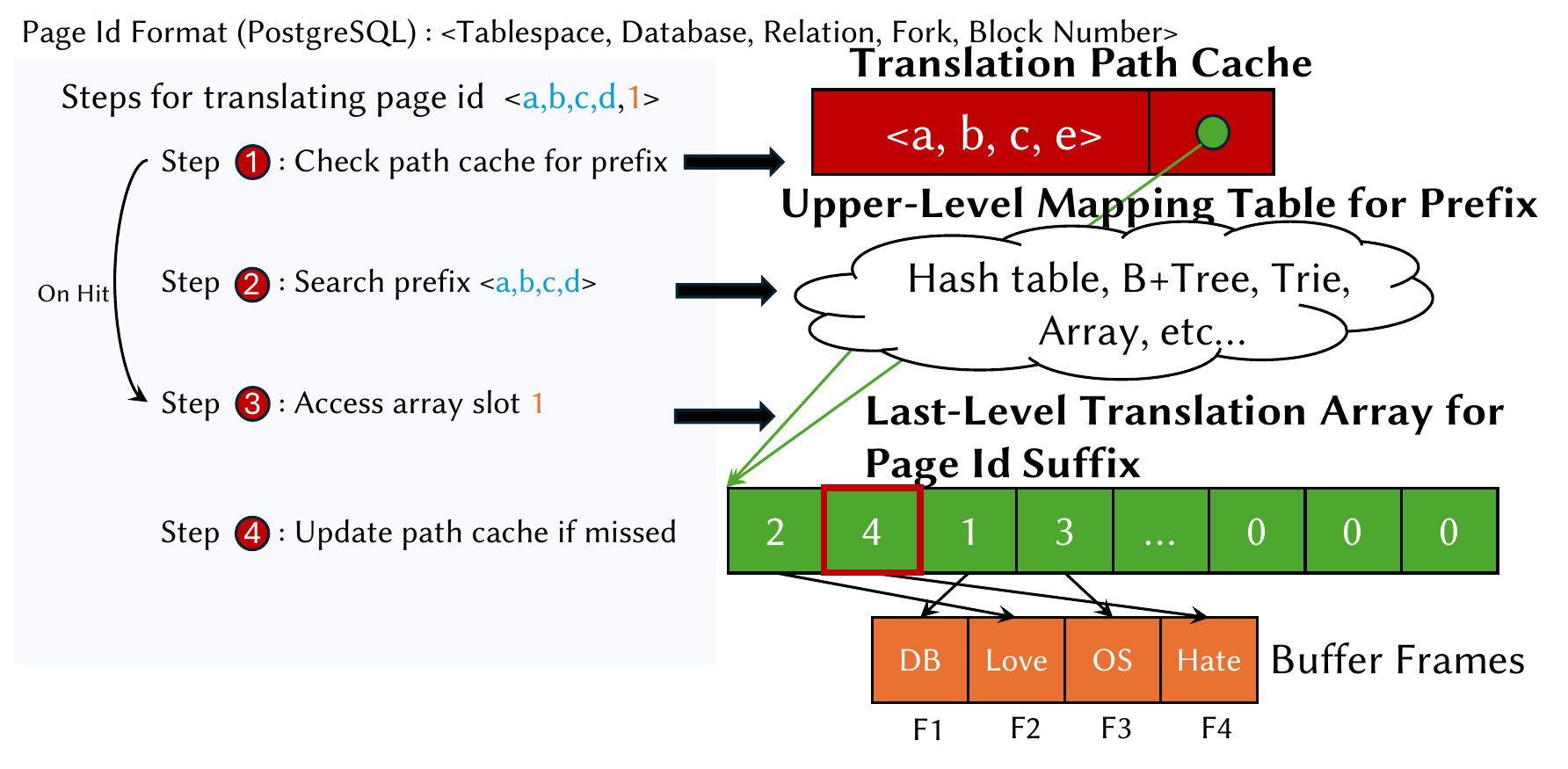}
    \Description{Step-by-step hierarchical translation workflow with translation-path caching. A page identifier is split into a prefix and suffix. Translation first checks the path cache; on a miss, the prefix is resolved through an upper-level search structure to locate the last-level translation array; then the suffix directly indexes that array. The resolved mapping is inserted into the path cache for reuse.}
    \caption{\captionTitle{Step-by-Step Hierarchical Translation with Path Caching}
    \calico decomposes each page identifier into a \emph{prefix} and a \emph{suffix}. The figure walks
    through four steps: (1) check the translation path cache; (2) on a miss, resolve the prefix through
    an upper-level index (e.g., radix tree, hash table, B$^+$-tree, or trie) to obtain a last-level
    translation array; (3) use the suffix to directly index that array on the hot path; and (4) update
    the path cache with the resolved prefix-to-array mapping.
    }
    \label{fig:multi-level-calico}
\end{figure}

\section{\calico DESIGN}
\label{sec:design}

While our evaluation in \cref{sec:translation} showed that array translation is promising,
the practical difficulty is managing sparse hierarchical PID spaces in real DBMSs: a naive flat array is
impractical for PostgreSQL/MySQL-scale identifier domains. We therefore start by showing how \calico
makes array translation practical with a sparse multi-level organization and translation-path caching,
then present the single-level entry format, memory-management mechanisms, and buffer-pool algorithms
that preserve the array fast path.

\subsection{Design Overview}
\label{sec:arch-overview}
Section~\ref{sec:array-takeaway} showed that array translation is performance-competitive, but a
deployable design must satisfy four constraints simultaneously: (1) keep array-indexed lookup on
the hot path, (2) handle sparse hierarchical PID spaces without allocating a monolithic flat array,
(3) preserve DBMS-managed fine-grained eviction while keeping frame memory huge-page-backed for TLB
efficiency, and (4) provide concurrency-safe access under pin, unpin, fault, and eviction races.
\calico addresses these constraints by separating logical translation control from physical memory
backing (\cref{fig:CalicoArch}). \cref{fig:CalicoArch} provides the architecture view, while
\cref{fig:multi-level-calico} provides a step walkthrough of multi-level translation with path caching.

\parhead{Hierarchical translation arrays}
\calico stores logical translation state in DBMS-managed structures where upper levels map PID
prefixes to lower-level arrays, and the last-level array remains array-indexed on the hot path. The
upper-level mapping component is explicit in \cref{fig:CalicoArch} and instantiated as an index over
prefixes in \cref{fig:multi-level-calico}.

\parhead{Huge-page-backed frame memory}
\calico stores page contents in a contiguous frame region backed by OS huge pages to reduce TLB
pressure. Frame-memory mappings remain stable across normal page eviction and reload.

\parhead{Hole-punching array}
\calico maintains a lightweight reference-count array over translation-entry groups for each translation array so fully cold
regions can be reclaimed and returned to the OS.

\subsection{Hierarchical Translation and Path Caching}
\label{sec:multi-level-calico}
To tackle the problem of large sparse PID spaces, \calico introduces hierarchical translation so memory is
proportional to active regions rather than the full logical PID domain. Upper levels map PID
prefixes to lower-level translation arrays, and those lower-level arrays are allocated only when a
prefix becomes active. Prefixes with no resident pages keep no materialized last-level array, so
empty PID regions do not consume translation memory. The number of levels is not fixed; it is chosen
based on PID-space sparsity/structure of the target DBMS.

While multi-level translation introduces extra indirection, the key observation is that modern
DBMSs organize page identifiers hierarchically, which induces strong locality in prefix components.
B-tree traversals, HNSW traversals, and scans repeatedly access pages within the same
relation/index region~\cite{sqlserver-index-storage,oracle-index-storage,postgres-index-storage,postgres-btree,mysql-btree,pgvector-hnsw}. \calico exploits this locality with translation-path caching.
\cref{fig:multi-level-calico} walks through translation of page identifier $p=(prefix,suffix)$:
first check the path cache, then resolve the prefix through the upper-level index on a miss, then
perform suffix-based array indexing in the resolved last-level array, and finally update the cache
with the resolved prefix-to-array mapping. When subsequent accesses reuse the same prefix, upper-level
traversal is bypassed and only the final array lookup remains on the hot path.

A practical decomposition follows storage hierarchy. The \emph{prefix} identifies a stable container
region (for example, database/table/index or relation/fork), and the \emph{suffix} identifies the
page within that region. In most DBMS layouts, the suffix is the page or block number within a
table/index, which keeps the last-level translation array densely indexable by array offset.

Take PostgreSQL as a concrete instantiation. For PID
\texttt{<Tablespace, Database, Relation, Fork, Block>}, \calico uses
\texttt{prefix = <Tablespace, Database, Relation, Fork>} to select the last-level translation array
and \texttt{suffix = <Block Number>} for direct indexing in that array
(\cref{fig:multi-level-calico}). This matches observed locality where accesses repeatedly remain in
the same relation/fork region while block numbers vary. The same principle generalizes to MySQL's
hierarchical \texttt{<space\_id, page\_no>} identifiers, where \texttt{space\_id} is a natural
prefix and \texttt{page\_no} is a natural suffix. Additional levels can be added when PID-space
sparsity or locality requires them.

\subsection{On-demand Array Memory Management}
After upper-level prefix resolution, every access goes through a last-level translation array. This
hot path must satisfy three requirements simultaneously: (1) fast array-indexed translation,
(2) concurrency control for reads/updates/eviction, and (3) memory usage proportional to the
working set despite sparse logical PID spaces.

\calico encodes each last-level \texttt{TranslationEntry} as one 64-bit atomic word. The
\textbf{latch state} (8 bits) supports exclusive/shared synchronization, the \textbf{version number}
(24 bits) enables optimistic validation for lock-free reads on the buffer frame, and the \textbf{frame ID} (32 bits)
identifies the buffer frame. A single load returns both translation and concurrency metadata.
Page access on the fast path is:
\begin{code}
TranslationEntry* te = &TranslationTable[pageId]; 
Frame* frame = frameMem + te->getFrameId();
\end{code}


\parhead{Memory efficiency and reclamation}
\label{sec:lazy-allocation}
The flat translation array introduces a potential space overhead: for a 16TB SSD with 4KB
pages, the table requires 32GB of memory (4G entries × 8 bytes). \calico leverages the
OS's on-demand memory allocation mechanism combined with \textbf{active hole-punching} to reduce the
\textit{physical} memory consumption so that it is proportional only to the working set size.

\parhead{Zero-Page Copy-on-Write at Startup} The translation array is allocated via \texttt{mmap}. Windows provides analogous virtual-memory reservation APIs~\cite{windows-virtualalloc}. The reserved region is initially backed by a shared zero-filled OS physical page~\cite{lwn-zero-page-cow}. When a thread first reads an entry, the OS triggers a page fault and establishes a read-only mapping to the shared zero-filled physical page, setting the copy-on-write (COW) bit in the page table entry. This read fault does not allocate physical memory. Only when an entry is \textit{written} (e.g., during \texttt{calico\_page\_fault\_handler} updating \texttt{frameId}) does the OS allocate a physical page, copy the zero page content, and update the mapping. This on-demand allocation pattern naturally aligns with \calico's access pattern: entries for cold pages that are never loaded remain on the shared zero page indefinitely, consuming no physical memory. Thus, the virtual size of the translation array can be proportional to the logical page ID space, while its physical memory footprint tracks only the working set.

\parhead{Zero-Value Entry Design for Lazy Loading} To exploit this mechanism, \calico's \texttt{TranslationEntry} encoding is carefully designed such that an all-zero 64-bit value represents an \textit{evicted} page state. Specifically, when all 64 bits are zero: (1) the frame ID field is 0, interpreted as \texttt{INVALID\_FRAME}, indicating no buffer frame is allocated; (2) the latch state field is 0, representing \texttt{Evicted}; and (3) the version number is 0. This invariant ensures that at system startup, when the entire translation array is logically filled with zeros (via the shared zero page), every entry correctly represents an evicted page requiring a page fault on first access. Crucially, when a page is evicted (\texttt{calico\_evict\_victim}), \calico writes a zero value back to the entry, returning it to the invalid state. This allows further memory reclamation through active hole-punching.

\parhead{Hole-Punching Array} \label{sec:metadata-array} While lazy allocation via OS demand paging ensures translation arrays consume physical memory proportional to active entries, cold regions may remain physically allocated long after pages are evicted. To address this, \calico introduces the Hole-Punching Array(\texttt{HPArray}), a lightweight reference counting structure that tracks the number of valid (non-evicted) translation entries within each OS page of the translation array, enabling active memory reclamation.

The \texttt{HPArray} is a flat array of atomic 32-bit counters, where each counter tracks one \textbf{entry group}: consecutive translation entries fitting within a single OS page. For a translation array with $N$ entries, \texttt{HPArray} requires $\lceil N / \text{entries\_per\_OS\_page} \rceil$ counters. With 512M translation entries and 4KB pages, this requires 1M counters (4MB); with 2MB huge pages, only 2048 counters. The \texttt{HPArray} itself is allocated lazily via \texttt{mmap}, consuming physical memory only on first write. Each counter reserves one bit as a lock to coordinate hole-punching operations (\cref{sec:buffer-algorithms}).

\parhead{Worst-Case Space Overhead}
In a pathological "sparse" scenario where only a single data page is resident within a 512-page group (2MB region), \calico cannot hole-punch the corresponding 4KB page in the metadata array. Consequently, the metadata overhead for that single resident page rises to 4KB. However, this worst-case overhead is identical to standard OS paging and vmcache. In x86-64, a single active 4KB page within a 2MB virtual memory mandates the allocation of a 4KB hardware page table. Furthermore, \calico potentially outperforms the OS in metadata reclamation: while swapped-out OS pages leave non-zero swap entries that prevent page table reclamation, \calico's eviction explicitly zeros entries, allowing our hole-punching mechanism to reclaim metadata memory even for datasets that have been fully evicted but was accessed previously.
If pathological sparsity is a concern, an \emph{adaptive hybrid} approach can further reduce overhead without modifying the hot path: when a region's occupancy drops below a density threshold (e.g., $<$1\%) and it is infrequently accessed, its few live entries are migrated to a small fallback hash table, and the entire region is hole-punched. We leave a full implementation to future work as this complicates the concurrency control for the fallback structure and may not be necessary in practice given common workload locality.

\subsection{Buffer-pool algorithms}
\label{sec:buffer-algorithms}
\cref{algo:calico} presents \calico's core buffer-pool algorithms for pinning, unpinning, page
fault handling, and eviction. We assume a thread-local path cache that stores the most recent
\texttt{(prefix, lastLevelArrayPtr)} mapping for each thread.

\begin{algorithm}[t]
\small
\SetKwFunction{FGetTE}{GetTranslationEntry}
\SetKwFunction{FSplitPid}{split\_pid}
\SetKwFunction{FLookupLeaf}{lookup\_leaf\_table}
\SetKwFunction{FPageFault}{calico\_page\_fault\_handler}
\SetKwProg{Fn}{Function}{:}{}
\DontPrintSemicolon

\Fn{\textsc{GetTranslationEntry}(pageId)}{
    $(prefix, suffix) \gets$ \FSplitPid{pageId}\;
    \If{\texttt{TLSPathCache.prefix = prefix}}{
        $last\_level\_array \gets \texttt{TLSPathCache.lastLevelArrayPtr}$\;
    }
    \Else{
        $last\_level\_array \gets$ \FLookupLeaf{prefix}\;
        $\texttt{TLSPathCache} \gets (prefix, last\_level\_array)$\;
    }
    \Return $\&last\_level\_array[suffix]$\;
}

\BlankLine

\Fn{\textsc{calico\_pin\_exclusive}(pageId)}{
    \While{\textnormal{true}}{
        $te \gets$ \FGetTE{pageId}\;
        $old\_e \gets *te$ \tcp{Atomic load}
        \If{$old\_e.frameId = \textsc{INVALID\_FRAME}$}{
            \FPageFault{pageId, te}\;
            \textbf{continue}\;
        }
        \If{$old\_e.state = \textsc{Unlocked}$ \textnormal{\textbf{and}} $te.\text{CAS}(old\_e, 
        (old\_e.frameId, old\_e.version, \textsc{Locked}))$}{
            \Return $frameMem + old\_entry.frameId$\;
        }
    }
}

\BlankLine

\Fn{\textsc{calico\_unpin\_exclusive}(pageId)}{
    $te \gets$ \FGetTE{pageId}\;
    $te.\text{set\_unlocked\_bump\_version}()$ \tcp{Unlock and bump version}
}

\BlankLine

\Fn{\textsc{calico\_optimistic\_read}(pageId, read\_func)}{
    \While{\textnormal{true}}{
        $te \gets$ \FGetTE{pageId}\;
        $old\_entry \gets *te$ \tcp{Atomic load}
        \If{$old\_entry.frameId = \textsc{INVALID\_FRAME}$}{
            \FPageFault{pageId, te}\;
            \textbf{continue}\;
        }
        \If{$old\_entry.state = \textsc{Locked}$}{
            \textbf{continue} \tcp{Spin until unlocked}
        }
        $read\_func(frameMem + old\_entry.frameId)$\;
        $new\_entry \gets *te$\;
        \If{$old\_entry.version = new\_entry.version$ \textnormal{\textbf{and}} $old\_entry.frameId = new\_entry.frameId$}{
            \Return \tcp{Success}
        }
    }
}

\caption{\calico Buffer Pool Algorithms}
\label{algo:calico}
\end{algorithm}

\begin{algorithm}[t]
\small
\SetKwFunction{FEvict}{calico\_evict\_victim}
\SetKwFunction{FAllocFrame}{allocate\_frame}
\SetKwFunction{FIoRead}{io\_read\_page}
\SetKwFunction{FIncRef}{increment\_metadata\_refcount}
\SetKwFunction{FGroupIdx}{group\_index}
\SetKwProg{Fn}{Function}{:}{}
\DontPrintSemicolon

\Fn{\textsc{calico\_page\_fault\_handler}(pageId, te)}{
    \While{$\neg te.\text{try\_lock}()$}{continue} 
    \If{$te.frameId \neq \textsc{INVALID\_FRAME}$}{
        $te.\text{unlock}()$\;
        \Return\;
    }
    $frameId \gets$ \FAllocFrame{}\;
    \If{$frameId = \textsc{INVALID\_FRAME}$}{
        $frameId \gets$ \FEvict{}\;
    }
    \FIoRead{pageId, $frameMem + frameId$}\;
    \tcp{Atomically increment counter for this entry group}
    $groupIdx \gets$ \FGroupIdx{te}\;
    \FIncRef{$groupIdx$} \;
    $te.\text{set\_frame\_and\_unlock}(frameId)$\;
}

\caption{\calico Page Fault Handler}
\label{algo:calico-fault}
\end{algorithm}

\begin{algorithm}[t]
\small
\SetKwFunction{FSelectVictim}{select\_victim\_page}
\SetKwFunction{FWriteBack}{write\_back\_if\_dirty}
\SetKwFunction{FHolePunch}{hole\_punch\_os\_page}
\SetKwFunction{FGroupBase}{group\_base\_addr}
\SetKwFunction{FGetMeta}{HPArray}
\SetKwProg{Fn}{Function}{:}{}
\DontPrintSemicolon

\Fn{\textsc{calico\_evict\_victim}()}{
    $victimId \gets$ \FSelectVictim{} \tcp{CLOCK, LRU, etc.}
    $te \gets$ \FGetTE{victimId}\;
    \While{$\neg te.\text{try\_lock}()$}{continue}
    $frameId \gets te.frameId$\;
    \FWriteBack{$frameMem + frameId$}\;
    $te.frameId \gets \textsc{INVALID\_FRAME}$ \tcp{Zero out frame id}
    \BlankLine
    \tcp{Atomically decrement MetadataArray refcount}
    $groupIdx \gets$ \FGroupIdx{te}\;
    $count \gets$ \FGetMeta{victimId}$[groupIdx].\textsc{lock\_and\_dec}()$\;
    $te.\text{unlock_evicted}()$ \tcp{Now zero out latch state}
    \If{$count = 0$}{ \tcp{Last valid entry evicted, reclaim memory}
        madvise$(\FGroupBase{te}, 4096, MADV\_DONTNEED)$
    }
    \FGetMeta{victimId}$[groupIdx].\textsc{unlock}()$\;
    \Return $frameId$\;
}

\caption{\calico Eviction with Hole-Punching}
\label{algo:calico-evict}
\end{algorithm}

\parhead{Exclusive Pin (\texttt{calico\_pin\_exclusive})} To modify a page, a thread must acquire an exclusive lock. The algorithm atomically loads the \texttt{TranslationEntry}, checks if the page is resident (valid frame ID), and uses compare-and-swap (CAS) to transition from \texttt{Unlocked} to \texttt{Locked}. If the page is not resident (\texttt{INVALID\_FRAME}), it triggers the page fault handler. On success, the thread receives a pointer to the frame holding the page data. Note that shared pins can be implemented similarly by storing the number of readers in the latch state of \texttt{TranslationEntry}.

\parhead{Exclusive Unpin (\texttt{calico\_unpin\_exclusive})} Releasing an exclusive pin is a single atomic operation: set the lock state to \texttt{Unlocked} and increment the version number. The version bump invalidates any concurrent optimistic readers that observed the old version, ensuring linearizability.

\parhead{Optimistic Read (\texttt{calico\_optimistic\_read})}\label{sec:optimistic-read-interface}
For read-heavy code paths, \calico exposes a lock-free optimistic read interface. The reader
snapshots a \texttt{TranslationEntry}, executes the read function on the target frame, and then
validates that the entry version and frame ID are unchanged and that the entry is not locked. This
avoids atomic pin/unpin traffic on the read path while preserving correctness under concurrent
eviction and frame modifications.

\parhead{Page Fault Handler (\cref{algo:calico-fault})} When a page is not resident, the fault handler acquires an exclusive lock on the translation entry and double-checks the frame ID (another thread may have already loaded the page). If still invalid, it allocates a frame (from the free list or via eviction), issues I/O to load the page data, then atomically increments the reference count in the \texttt{MetadataArray} for the OS page group containing this translation entry. Finally, it updates the entry with the new frame ID and releases the lock. The translation entry lock prevents duplicate I/O when multiple threads fault on the same page, while incrementing the metadata counter before publishing the frame ID ensures the group cannot be hole-punched during page fault.

\parhead{Eviction (\cref{algo:calico-evict})} \calico uses a standard replacement policy (CLOCK) to select a victim page. The algorithm acquires an exclusive lock on the victim's translation entry, writes back the frame if dirty, and invalidates the entry by setting the frame ID to \texttt{INVALID\_FRAME}. Next, it atomically locks the metadata counter and decrements the reference count for the corresponding OS page group. Crucially, the translation entry is unlocked only \emph{after} acquiring the metadata lock. If the count reaches zero (all entries in the group evicted) after decrement, the thread performs hole-punching via passing \texttt{MADV\_DONTNEED} hint to OS while still holding the metadata lock, then releases the lock. This ordering ensures that concurrent page faults must wait for the metadata lock before incrementing the counter, preventing any thread from installing a new frame ID in a page being hole-punched.

\subsection{Discussion}

Several index structures employ array-based mapping tables as an
\emph{internal} indirection layer. For example, Bw-tree~\cite{Levandoski2013,wang2018building}
and Bf-tree~\cite{hao2024bf} map logical node id to
in-memory pointers or offsets on flash storage in order to support latch-free operations and track node locations. \calico differs
from these works in two ways. First, their mapping
tables live inside a single index implementation, whereas \calico's
translation array is a \emph{global} buffer-pool structure shared
across all indexes. Therefore, the improvements impact all indexes built on the buffer manager. Second, Bw-tree/Bf-tree mappings must
store non-zero pointers or offsets for every live node, so they
cannot exploit the "all-zero = evicted" invariant, zero-page
sharing, or active hole-punching that \calico employs.

\section{IMPLEMENTATION AND OPTIMIZATIONS}
\label{sec:implementation}

This section presents \calico's group prefetch interface and then describes system integration in PostgreSQL and \texttt{pgvector}.

\subsection{Group prefetch}

Modern workloads like graph-based vector search exhibit predictable access patterns: future page accesses are often known in advance. In proximity-graph-based vector search~\cite{malkov2018efficient, fu2017fast, jayaram2019diskann}, when visiting a graph node, the algorithm must probe neighboring nodes whose IDs are stored in the current node. \calico exploits this predictability through a \textbf{group prefetch interface} that issues parallel reads to the translation array and prefetches buffer frame memory. This parallelism operates at two levels:

\begin{itemize}[leftmargin=*]
    \item \textbf{Memory-Level Parallelism}: For resident pages, parallel translation array reads allow the CPU to issue multiple independent loads simultaneously. Modern CPUs can issue  parallel memory loads~\cite{mahling2025fetch,mlp-m2-m4, mlp-intel-arm}, hiding memory latency.
    \item \textbf{I/O-Level Parallelism}: For non-resident pages, knowing multiple page IDs in advance enables batched asynchronous I/O, saturating storage bandwidth and reducing total latency.
\end{itemize}

\cref{algo:prefetch} presents the group prefetch algorithm. The interface accepts page IDs with corresponding in-page offsets for targeted prefetching. The algorithm operates in three phases: (1) issues prefetch instructions for translation entries to bring them into cache; (2) issues prefetch instructions for resident page frames at specified offsets while collecting non-resident page IDs; (3) submits batched reads for non-resident pages. This exposes batching semantics, enabling indexes to exploit modern hardware capabilities.

\subsection{PostgreSQL integration}

We implement multi-level \calico in PostgreSQL v18, which uses a 5-level hierarchical page identifier BufferTag~\cite{pg-buffertag}. We use the last-level 32-bit BlockNumber as the low bits and the remaining fields as the high bits identifying the relation. Our implementation uses a hash table for top-level mapping and flat arrays for last-level translation arrays. We extended PostgreSQL's buffer manager with \calico's group prefetch and optimistic read interfaces. The implementation adds approximately 2.6K lines of C code. The modifications are mostly isolated within the buffer manager, preserving compatibility with existing components.

\parhead{Translation Path Caching via Thread-Local Storage} To reduce repeated top-level lookups, we leverage \texttt{thread\_local} storage to cache the most recently accessed translation path. Each thread maintains a cache entry containing: (1) the last accessed \texttt{BufferTag}, and (2) the corresponding last-level translation array pointer. On each page access through the buffer manager, the thread first checks if the current \texttt{BufferTag}'s high bits matches the cached value. On a cache hit, the thread directly uses the cached last-level array pointer. Note that this optimization can be implemented in any buffer manager in database system with hierarchical page IDs and thread-local storage, without coupling with other components (e.g., indexing layer, query execution) in the systems.

\subsection{pgvector integration}

We then modified the \texttt{pgvector} extension (about 600 lines) to leverage these interfaces during HNSW graph traversal. PostgreSQL's existing page access requires complex pin/unpin operations using atomic writes that limit memory-level parallelism~\cite{david2013everything,mckenney2010memory}. By using \calico's optimistic read interface (\cref{sec:optimistic-read-interface}), pgvector bypasses these atomic operations: when visiting a graph node, pgvector extracts neighbor page IDs and prefetch neighbors, then uses \texttt{calico\_optimistic\_read} to access neighbor data without pin/unpin overhead. When validation fails, we revert to standard pin/unpin operations to avoid repeated wasted work. This combination enables memory-level parallelism for resident pages and I/O-level parallelism for non-resident pages. 

\begin{algorithm}[t]
\small
\SetKwFunction{FPrefetchPages}{calico\_read\_pages}
\SetKwFunction{FGetTE}{GetTranslationEntry}
\SetKwProg{Fn}{Function}{:}{}
\DontPrintSemicolon

\Fn{\textsc{calico\_prefetch\_group}(pids, offsets)}{
    $non\_resident\_pids \gets {[\,]}$\;
    \tcp{Prefetch translation entries}
    \ForEach{$pageId \in pids$}{
        $te \gets$ \FGetTE{pageId}\;
        \text{prefetch}(te)\;
    }
    \BlankLine
    \tcp{Prefetch resident pages and collect non-resident ones}
    \ForEach{$i \in [0, \text{length}(pids))$}{
        $pageId \gets pids[i]$\;
        $offset \gets offsets[i]$\;
        $te \gets$ \FGetTE{pageId}\;
        $entry \gets *te$\;
        \If{$entry.frameId \neq \textsc{INVALID\_FRAME}$}{
             \text{prefetch}(\text{FrameMemory}[entry.frameId] + offset)\;
        }
        \Else{
            $non\_resident\_pids.\text{append}(pageId)$\;
        }
    }
    \BlankLine
    \If{$non\_resident\_pids \neq \emptyset$}{
        \FPrefetchPages{$non\_resident\_pids$}\;
    }
}

\caption{\calico Group Prefetch Algorithm}
\label{algo:prefetch}
\end{algorithm}

\section{EXPERIMENTAL EVALUATION}
\label{sec:eval}
Our evaluation validates \calico's performance and scalability across modern workloads. We aim 
to evaluate:

\begin{enumerate}[leftmargin=*]
    \item Array-based translation overhead compared to mmap, pointer swizzling, hash tables on vector search and OLTP workloads.
    \item Memory overhead of array-based translation with hole-punching.
    \item End-to-end performance gains in PostgreSQL.
    \item Generality of \calico on different CPU platforms.
\end{enumerate}


Unless stated otherwise, all experiments are conducted on a dual-socket server with two
AMD EPYC 7513 processors (32 cores per socket, 64 cores/128 threads total, 2.6 GHz base frequency),
504 GB DDR4 memory, and a Samsung PM9A3 3.84TB NVMe SSD (1M random 4KB read IOPS). The system runs
CentOS Linux 8.5. We set the CPU governor to performance mode to ensure
consistent results.
We use Linux transparent huge
pages feature via \texttt{madvise(MADV\_HUGEPAGE)} hint to enable 2MB huge pages to back frame
memory, except vmcache which uses 4KB pages due to the eviction granularity problem. We use huge
pages to back hash table memory in hash table-based pools as well.

\subsection{Workloads and Baselines}
\label{sec:eval-workloads}

We evaluate \calico on two workloads:

\begin{itemize}[leftmargin=*]
   
    \item \textbf{Vector Search with HNSW}:
    We evaluate vector similarity search using HNSW indexing algorithm. The index builds on buffer manager interface for accessing pages on storage.
    Baselines include \textbf{vmcache}, \textbf{USearch}~\cite{usearch} (specialized in-memory HNSW
    library using \texttt{mmap}), and \textbf{hash table variants}.

     \item \textbf{OLTP Workloads with B+tree}:
    We use YCSB-C (read-heavy) and TPC-C (write-heavy)
    workloads, both operating on B+tree indexes. Baselines include \textbf{vmcache}~\cite{vmcache}
    (mmap-based buffer pool), \textbf{LMDB}~\cite{henry2019howard} v0.9.31 (a mmap-based key-value store), \textbf{WiredTiger} v10.0.2 (hash table), and
    \textbf{LeanStore}~\cite{leis2018leanstore} at commit 629b41a (pointer swizzling) . We use
    the lowest isolation level for LeanStore and WiredTiger and disable write-ahead-logging to focus on buffer management performance.

\end{itemize}


\subsection{Vector Search Workloads}

We evaluate \calico on HNSW-based vector search using two datasets shown in \cref{tab:vector-datasets}. 
We implement an in-process vector search library~\cite{caliby} with HNSW algorithm~\cite{malkov2018efficient} and evaluated it on different buffer managers:
\begin{itemize}[leftmargin=*]
    \item \textbf{Calico}: direct array translation
    \item \textbf{vmcache}: virtual memory-based translation
    \item \textbf{Hash table variants}: \texttt{std::unordered}, \texttt{absl::flat\_hash}, Lock-Free Linear Probing Hash Map
    \item \textbf{USearch}~\cite{usearch}: specialized in-memory HNSW library using \texttt{mmap}
\end{itemize}
All experiments run at 64 threads with the same HNSW search parameters (M=16, ef\_construction=100, ef\_search=50). We enable optimistic reads and group prefetch for all buffer managers. For all hash table-based pools except the lock free hash table, we partition the hash table and use dedicated lock per partition to reduce lock contention when accessing the hash tables.

\begin{table}[t]
    \centering
    \small
    \begin{tabular}{lrr}
        \toprule
        \textbf{Dataset} & \textbf{Vectors} & \textbf{Dimension} \\
        \midrule
        DEEP~\cite{deep10M} & 10 million & 96 \\
        SIFT~\cite{sift10M} & 10 million & 128 \\
        \bottomrule
    \end{tabular}
    \caption{Vector search datasets used for evaluation.}
    \label{tab:vector-datasets}
\end{table}

\begin{figure}[t]
    \includegraphics[width=0.5\textwidth]{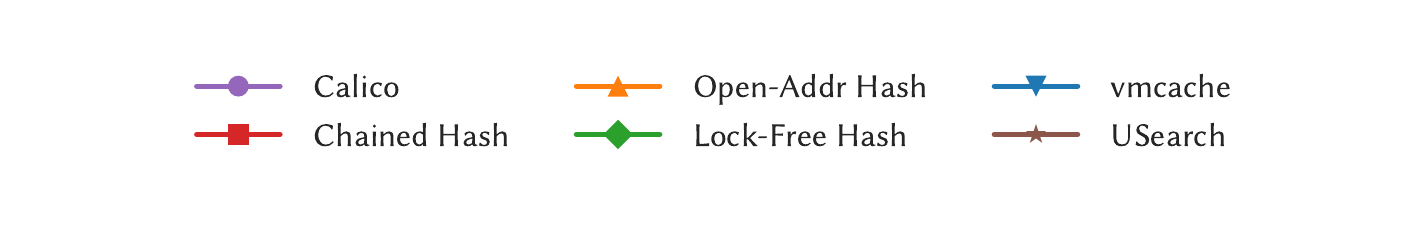}
    \begin{subfigure}{0.23\textwidth}
        \includegraphics[width=\textwidth]{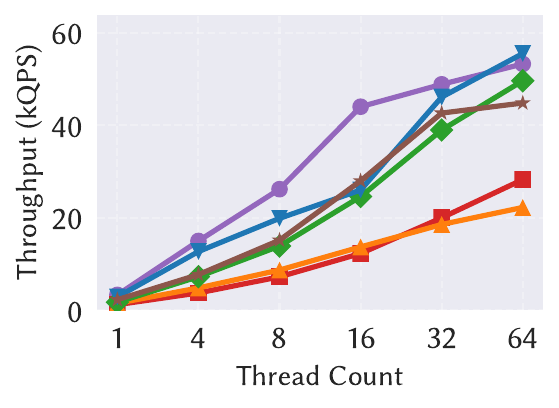}
        \subcaption{\textbf{DEEP10M}}
        \label{fig:deep10m-scalability}
    \end{subfigure}
    \hfill
    \begin{subfigure}{0.23\textwidth}
        \includegraphics[width=\textwidth]{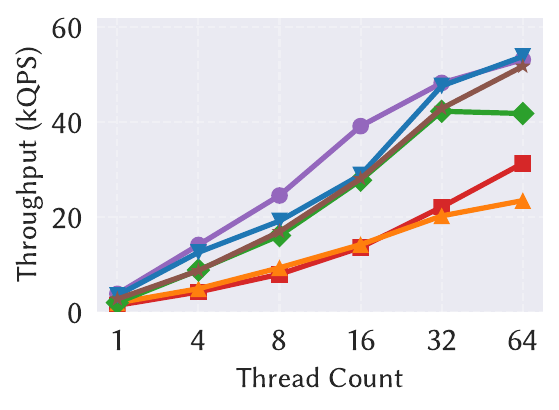}
        \subcaption{\textbf{SIFT10M}}
        \label{fig:sift10m-scalability}
    \end{subfigure}
    \Description{Two line charts show in-memory HNSW throughput scaling with thread count for DEEP10M and SIFT10M. Calico and vmcache are the top performers and scale similarly, while hash-table variants trail, especially at higher thread counts.}
    \caption{
        \captionTitle{Vector Search (In-Memory)}
        Measured throughput of the HNSW index on DEEP10M and SIFT10M when the entire data set fits
        in memory using 1--64 threads.
    }
    \label{fig:hnsw-inmem-scalability}
\end{figure}

\parhead{In-Memory}
\cref{fig:deep10m-scalability,fig:sift10m-scalability} show throughput scaling when the working set
fits entirely in memory. With a single thread, \calico leads with 3.3K QPS on DEEP10M and
3.8K QPS on SIFT10M, outperforming vmcache (2.9K and 3.5K QPS), USearch (2.3K and 2.7K
QPS), Lock-Free Hash (1.8K and 1.9K QPS), and Chained Hash (1.2K and 1.4K QPS). At 64 threads,
\calico achieves 53.2K QPS on DEEP10M and 53.1K QPS on SIFT10M, matching vmcache's
55.5K QPS and 53.8K QPS respectively, demonstrating that \calico's performance
scales well while maintaining its single-thread advantage. The hash-based designs collapse under
contention: Chained Hash achieves only 28.3K QPS and 31.3K QPS (53-59\% of
\calico), while Lock-Free Hash reaches 49.6K QPS and 41.8K QPS (93\% and 79\% of
\calico). These results validate that array translation matches \texttt{mmap} performance
in-memory.

\begin{figure}[t]
    \includegraphics[width=0.5\textwidth]{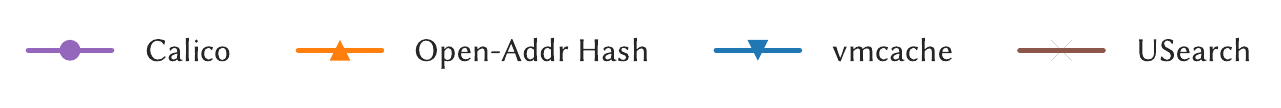}
    \begin{subfigure}{0.23\textwidth}
        \includegraphics[width=\textwidth]{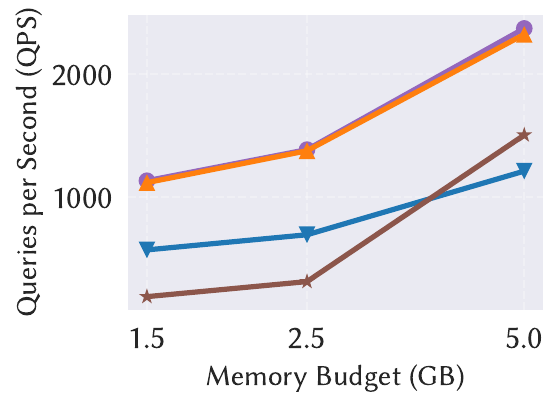}
        \subcaption{\textbf{DEEP10M}}
        \label{fig:oom-deep10m}
    \end{subfigure}
    \hfill
    \begin{subfigure}{0.23\textwidth}
        \includegraphics[width=\textwidth]{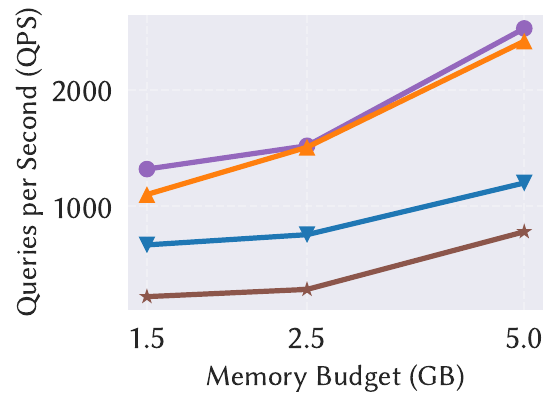}
        \subcaption{\textbf{SIFT10M}}
        \label{fig:oom-sift10m}
    \end{subfigure}
    \Description{Two line charts show larger-than-memory HNSW throughput as memory budget changes for DEEP10M and SIFT10M. Calico stays above vmcache and far above USearch as memory pressure increases.}
    \caption{
        \captionTitle{Vector Search (Larger-than-Memory)}
        Throughput comparison across buffer managers with varying memory budgets at 64 threads.
        \calico maintains superior performance, outperforming vmcache and USearch under memory 
        pressure by 2.11$\times$ and 5.99$\times$, respectively.
    }
    \label{fig:hnsw-oom}
\end{figure}

\parhead{Larger-than-Memory}
The results in \cref{fig:hnsw-oom} show that when working sets exceed available memory, \calico
outperforms both vmcache and USearch by 2--6$\times$. The
performance gap stems from fundamental differences in I/O management. USearch relies on
\texttt{mmap} with synchronous page faults: when data is not resident, the OS blocks threads until
I/O completes, serializing execution. At 1.5GB memory, USearch only achieves 189-222 QPS
(6$\times$ worse than \calico). All other buffer managers (\calico, vmcache, and hash tables) use
prefetch interfaces to issue parallel I/O requests, avoiding blocking page faults. This explains
USearch's steep degradation under memory pressure.

Among prefetch-enabled systems, \calico outperforms vmcache by 2$\times$ across
memory budgets (e.g., 2.37K vs 1.21K QPS at 5GB on DEEP10M). The root cause for this difference is 
TLB shootdown overhead. When vmcache evicts pages via \texttt{madvise(MADV\_DONTNEED)}, the OS 
interrupts all 64 threads to invalidate TLB entries, which is a synchronization bottleneck that 
serializes execution. \calico avoids this problem entirely because it evicts pages in 
user space without OS involvement, eliminating TLB shootdowns. This architectural advantage 
is critical when the system incurs memory pressure where frequent evictions amplify the cost 
difference.

\begin{figure}[t]
     \includegraphics[scale=0.4]{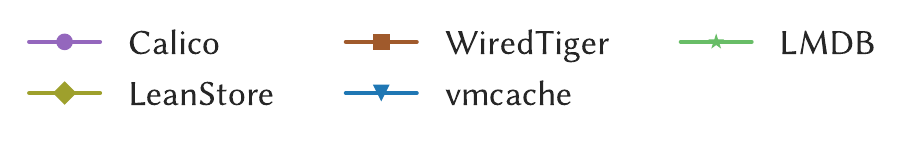}
     \centering
        \begin{subfigure}{0.21\textwidth}
         \includegraphics[scale=0.45]{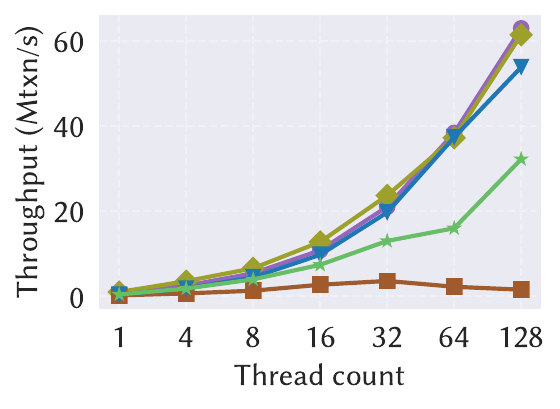}
        \subcaption{YCSB-C}
        \label{fig:btree-threads-ycsbc}
     \end{subfigure}
     \hfill
    \centering
        \begin{subfigure}{0.21\textwidth}
         \includegraphics[scale=0.45]{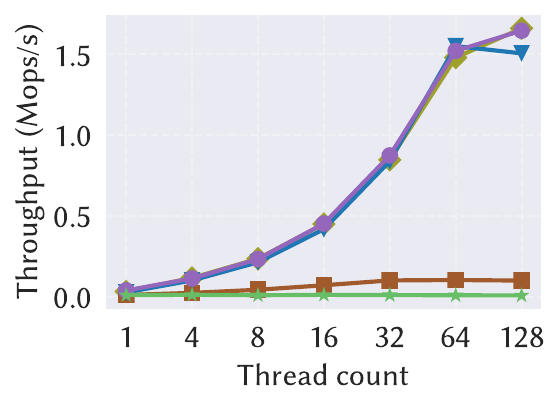}
        \subcaption{TPC-C}
        \label{fig:btree-threads-tpcc}
     \end{subfigure}
    \Description{Two plots show in-memory OLTP throughput scaling with thread count for YCSB-C and TPC-C. Calico, vmcache, and LeanStore are close, while WiredTiger and LMDB are substantially slower.}
    \caption{
        \captionTitle{In-memory YCSB-C/TPC-C Throughput Scaling}
        YCSB-C: 100 M entries=20 GB; TPC-C: 200 warehouses=40 GB;128 GB Buffer Pool }
     \label{fig:btree-inmem}
\end{figure}
\begin{figure}[h]
          \includegraphics[scale=0.4]{exps/oltp/btree_inmem_legend.pdf}

     \begin{subfigure}{0.23\textwidth}
        \includegraphics[scale=0.5]{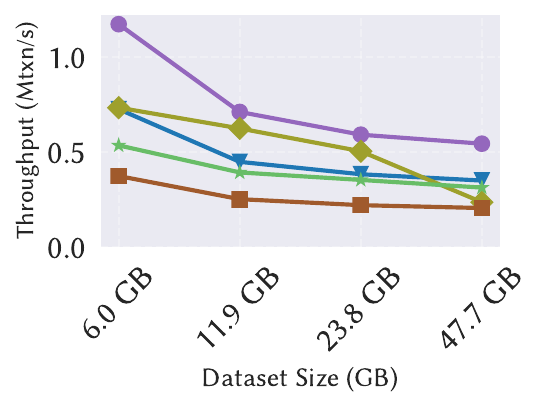}
        \subcaption{YCSB-C: Throughput}
        \label{fig:btree-oom-abs}
     \end{subfigure}
     \hfill
     \begin{subfigure}{0.23\textwidth}
        \includegraphics[scale=0.5]{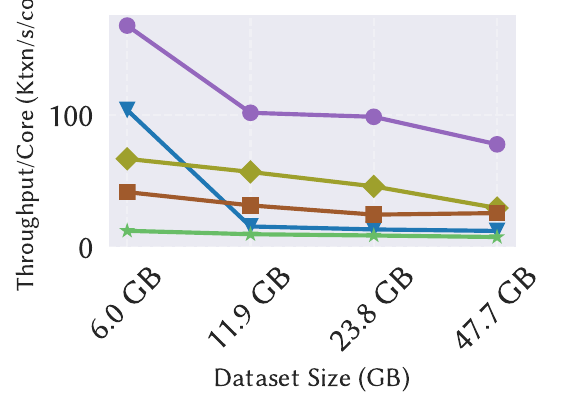}
        \subcaption{YCSB-C: Per-core Throughput}
        \label{fig:btree-oom-percore}
     \end{subfigure}
     
     \begin{subfigure}{0.23\textwidth}
        \includegraphics[scale=0.5]{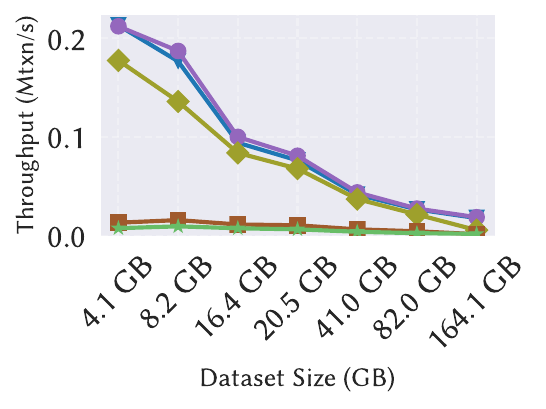}
        \subcaption{TPC-C: Throughput}
        \label{fig:tpcc-oom-abs}
     \end{subfigure}
     \hfill
     \begin{subfigure}{0.23\textwidth}
        \includegraphics[scale=0.5]{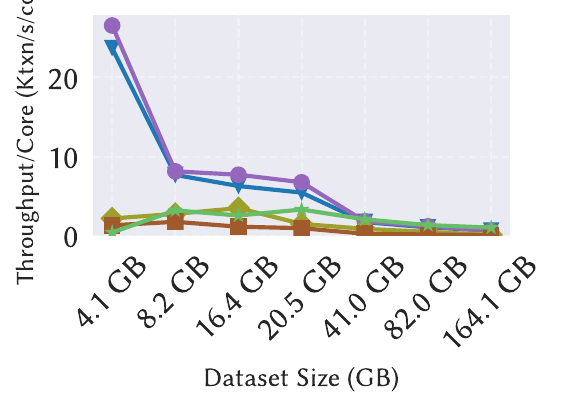}
        \subcaption{TPC-C: Per-core Throughput}
        \label{fig:tpcc-oom-percore}
     \end{subfigure}
    \Description{Four plots show larger-than-memory OLTP throughput and per-core throughput for YCSB-C and TPC-C. Calico generally exceeds vmcache and matches or exceeds LeanStore as datasets grow beyond buffer-pool capacity.}
    \caption{
        \captionTitle{OLTP Workloads (Larger-than-Memory)}
        Measured throughput for YCSB-C (4GB buffer pool) and TPC-C (8GB buffer pool).
    }
    \label{fig:btree-oom}
\end{figure}

\parhead{Summary} In-memory, \calico 
matches vmcache and USearch. Out-of-memory, \calico's explicit I/O control delivers 2$\times$ higher throughput 
than vmcache  and 6$\times$ higher than USearch.

\subsection{OLTP Workloads}
We evaluate \calico on YCSB-C and TPC-C using B+tree index in two scenarios: \textbf{in-memory} and \textbf{larger-than-memory}.

\parhead{In-Memory}
\cref{fig:btree-inmem} shows throughput scaling from 1 to 128 threads. For YCSB-C (\cref{fig:btree-threads-ycsbc}), LeanStore, vmcache, and \calico exhibit similar scalability as all three use optimistic reads. At single-thread, LeanStore (1.02M ops/s) outperforms \calico (935K ops/s) by only 9\% despite using huge pages for buffer frames and eliminating indirection via pointer swizzling. LMDB lags behind while WiredTiger is significantly slower and does not scale.

For TPC-C (\cref{fig:btree-threads-tpcc}), write contention differentiates systems. At 128 threads, LeanStore (1.66M txn/s), \calico (1.65M txn/s), and vmcache (1.51M txn/s) are comparable, while WiredTiger (99K txn/s) and LMDB (7.4K txn/s) are significantly slower. Note that LMDB by design only allows one writer and perform out-of-place writes which severely limits its scalability under write-heavy workload such as TPC-C. WiredTiger suffers from latch contention under high write workloads. \calico matches LeanStore's performance without complex engineering effort, demonstrating that array-based translation can rival pointer swizzling.

\parhead{Larger-Than-Memory}
\cref{fig:btree-oom} compares systems as dataset size exceeds buffer pool capacity (4GB for YCSB-C, 8GB for TPC-C) with 64 threads.

For YCSB-C (\cref{fig:btree-oom-abs,fig:btree-oom-percore}), \calico (1.17M txns/s at 6GB) outperforms vmcache/LeanStore by 1.6$\times$. As dataset grows to 47.7GB, \calico sustains 543K txns/s while vmcache and LMDB degrade to 349K and 312K txns/s respectively. Their TLB-shootdown overhead during page eviction causes high CPU consumption, resulting in low per-core throughput. WiredTiger's throughput is significantly lower due to 32KB page size.

For TPC-C (\cref{fig:tpcc-oom-abs,fig:tpcc-oom-percore}), \calico, vmcache, and LeanStore perform comparably in terms of absolute throughput and significantly outperforming LMDB and WiredTiger. However, LeanStore has much lower throughput per core compared to \calico/vmcache because its transaction worker threads spin waits for free frames to be produced by page provider threads that write dirty pages back to storage. In contrast, \calico/vmcache perform eviction in the transaction worker threads. LMDB's throughput is severely limited by its single-writer design under write-heavy workloads.

\begin{figure}[t]
     \begin{subfigure}{0.21\textwidth}
        \includegraphics[scale=0.42]{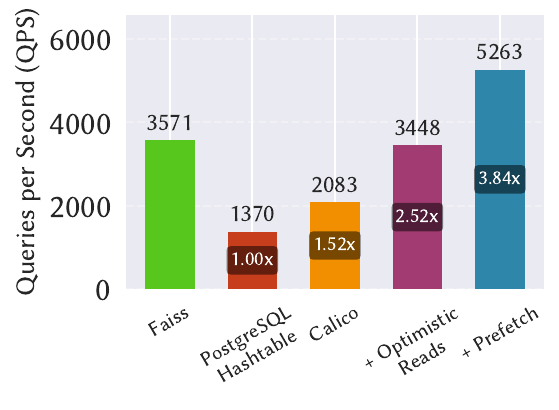}
        \subcaption{SIFT10M, 32GB Pool}
        \label{fig:pgvector-inmem-sift10m}
     \end{subfigure}
     \hfill
     \begin{subfigure}{0.21\textwidth}
        \includegraphics[scale=0.42]{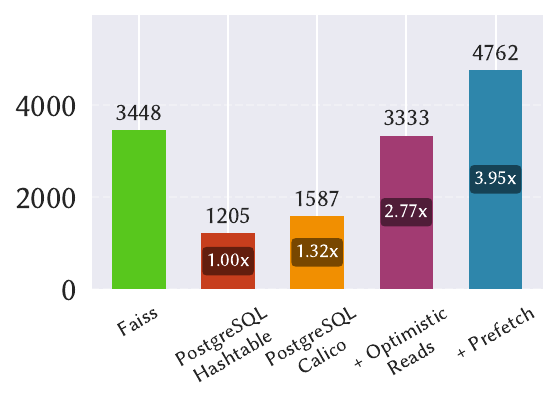}
        \subcaption{DEEP10M, 32GB Pool}
        \label{fig:pgvector-inmem-deep10m}
     \end{subfigure}
     
     
     \begin{subfigure}{0.21\textwidth}
        \includegraphics[scale=0.42]{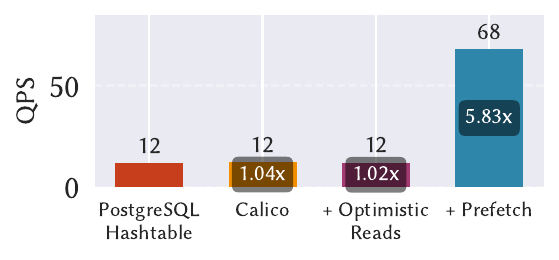}
        \subcaption{SIFT10M, 2GB Pool}
        \label{fig:pgvector-oom-sift10m}
     \end{subfigure}
     \hfill
     \begin{subfigure}{0.21\textwidth}
        \includegraphics[scale=0.42]{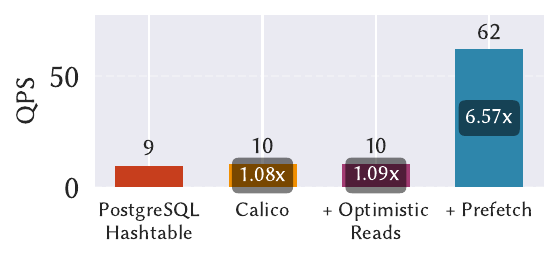}
        \subcaption{DEEP10M, 2GB Pool}
        \label{fig:pgvector-oom-deep10m}
     \end{subfigure}
    \Description{Four bar charts show PostgreSQL pgvector performance on SIFT10M and DEEP10M with 32GB and 2GB buffer pools. Calico improves both in-memory and larger-than-memory vector search throughput relative to PostgreSQL's baseline path.}
    \caption{
        \captionTitle{PostgreSQL Integration (Vector Search)}
        Top row: 32GB buffer pool (in-memory), \calico optimizations result in 2.9--3.95$\times$
        speedup, matching Faiss. Bottom row: 2GB buffer pool (larger-than-memory), group prefetch
        achieves 5.83--6.57$\times$ speedup.
    }
    \label{fig:pgvector-perf}
\end{figure}

\begin{figure}[t]
    \begin{subfigure}{0.25\textwidth}
        \centering
        \includegraphics[width=\linewidth]{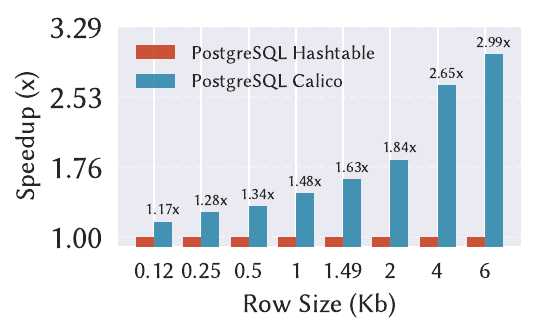}
        \subcaption{SUM query speedup across row sizes}
        \label{fig:postgres-seqscan-sum}
    \end{subfigure}
    \hfill
    \begin{subfigure}{0.215\textwidth}
        \centering
        \includegraphics[width=\linewidth]{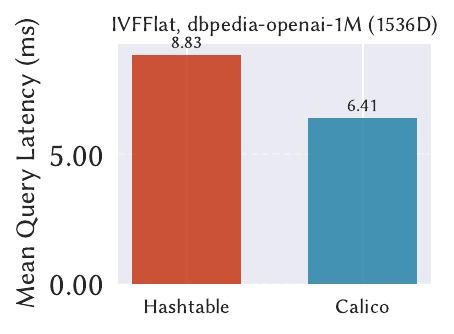}
        \subcaption{IVFFlat Query Latency}
        \label{fig:postgres-ivfflat-openai}
    \end{subfigure}
    \Description{Two bar charts. Left: PostgreSQL SUM sequential-scan speedup across row sizes with Hashtable normalized to 1.00x. Right: mean query latency for PostgreSQL Hashtable and Calico on DBPedia OpenAI-1M with IVFFlat (1536 dimensions, x1 duplication, 32GB buffer pool).}
    \caption{
        \captionTitle{PostgreSQL Integration (Sequential Scan)}
        Left: relative speedup of \calico over PostgreSQL Hashtable for \texttt{SUM} sequential
        scan, where Hashtable is normalized to 1.00$\times$. Right: IVFFlat query latency on
        DBPedia OpenAI-1M (1536D, 32GB), comparing PostgreSQL Hashtable and \calico.
    }
    \label{fig:postgres-seqscan}
\end{figure}

\subsection{PostgreSQL Integration}
\label{sec:eval-pg}
We next evaluate \calico's impact on PostgreSQL v18 using two representative workloads: (1)
vector similarity search via the HNSW index with \texttt{pgvector}, and (2) heap scans with varying
row sizes. We use a single PostgreSQL client in Python for the experiments.

\parhead{Vector Search}
\cref{fig:pgvector-perf} shows query throughput
for SIFT10M and DEEP10M datasets using \texttt{pgvector}. We start with vanilla PostgreSQL buffer manager
and incrementally enable array indexing, optimistic reads, and group prefetch. We additionally
compare with Faiss~\cite{Faiss} (in-memory-only vector search library). The cache hit rate
of the translation path caching is close to 100\% because PostgreSQL stores all pages of the HNSW
index within the same relation.

For SIFT10M (\cref{fig:pgvector-inmem-sift10m}), \calico array translation achieves
1.52$\times$ speedup over the PostgreSQL hashtable baseline. Adding optimistic reads
improves to 2.5$\times$ speedup by eliminating atomic pin/unpin overhead and cache
coherence traffic~\cite{david2013everything,mckenney2010memory}. Full \calico with group prefetch
reaches 3.84$\times$ speedup, demonstrating that workload-aware prefetching exploits
memory-level parallelism. For DEEP10M (\cref{fig:pgvector-inmem-deep10m}), the
improvements are similar: array translation achieves 1.32$\times$ speedup, optimistic reads
reach 2.77$\times$ speedup, and full \calico with group prefetch reaches
3.95$\times$ speedup.

\parhead{Larger-than-memory}
The bottom row of \cref{fig:pgvector-perf} shows
performance with a 2GB buffer pool, where the HNSW index significantly exceeds memory. In this
I/O-bound scenario, base \calico and optimistic reads provide modest improvements
(1.04--1.09$\times$) since I/O latency dominates. However, group prefetch achieves dramatic
speedups: 5.83$\times$ for SIFT10M and 6.57$\times$ for DEEP10M. \calico's
prefetch API effectively hides I/O latency by exploiting HNSW's graph structure to issue parallel
I/O requests.

\parhead{Sequential Scan}
\cref{fig:postgres-seqscan-sum} reports sequential \texttt{SELECT SUM(column)} scans on 15GB of
data with row sizes from 0.125 to 6 Kb (128 to 6144 bytes). The figure is normalized to PostgreSQL Hashtable
(1.00$\times$ baseline). \calico outperforms Hashtable at every row size, with speedup increasing
from 1.17$\times$ (128B) to 2.99$\times$ (6144B), and 1.70$\times$ geometric-mean speedup across
all row sizes. This trend matches the locality analysis in \cref{sec:translation}: hash-table
translation scatters consecutive page IDs across buckets, while \calico keeps adjacent translation
entries contiguous and preserves prefetch-friendly access. \cref{fig:postgres-ivfflat-openai} extends the comparison to IVFFlat on the DBPedia
OpenAI-1M dataset~\cite{dbpedia-openai-1m} (1536 dimensions) with a 32GB buffer pool. IVFFlat is a partitioned-based vector index that performs linear scan over candidate vectors. \
\calico reduces mean query latency by 1.38$\times$ at the same Recall@3 (0.9072).

\parhead{Summary}
The PostgreSQL integration demonstrates that \calico delivers substantial
performance gains with minimal engineering effort,
showing its array-based translation and optimizations benefits access patterns from graph
traversals in vector search to sequential scans in OLAP queries.

\subsection{Translation Memory Overhead}

We evaluate translation memory consumption using TPC-C (100 warehouses, 18GB--800GB), YCSB-C
(614GB), and YCSB-D (800GB, read-latest workload). We compare three approaches: \textbf{Calico} with
hole-punching, \textbf{Hash Table} (absl::flat\_hash), and \textbf{vmcache}. Buffer pool configurations are 16GB and 128GB for TPC-C,
16GB for YCSB-C and YCSB-D. We configure hash table capacity to scale with buffer pool frames: 
$N$×$17$×$2$ bytes for $N$ frames (17 bytes per entry, 2$\times$ capacity for 50\% load
factor). We account for \emph{all} memory used for translation state: for hash-table-based
approaches this includes the hash table itself, and for vmcache this includes the memory consumed
by OS page tables in addition to the state array used for tracking resident pages. For \calico, we include the translation array and any metadata for hole-punching. 

\begin{figure}[t]
    \includegraphics[scale=0.5]{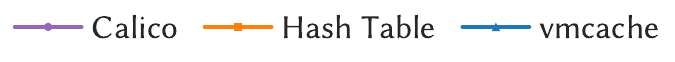}

     \begin{subfigure}{0.21\textwidth}
        \centering
        \includegraphics[scale=0.45]{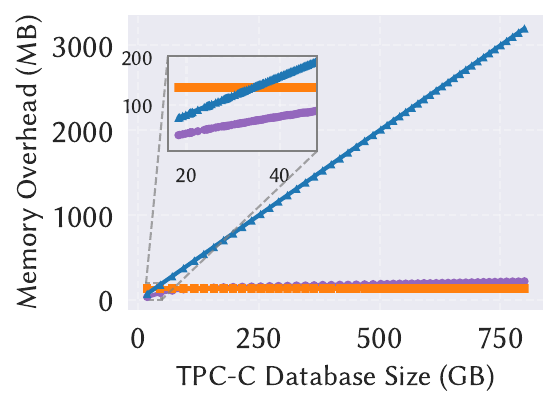}
        \subcaption{TPC-C, 16GB Pool}
        \label{fig:translation-space-overhead-16gb}
     \end{subfigure}
     \hfill
     \begin{subfigure}{0.215\textwidth}
        \centering
        \includegraphics[scale=0.45]{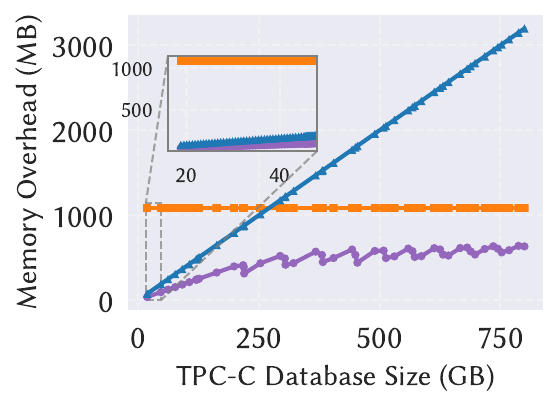}
        \subcaption{TPC-C, 128GB Pool}
        \label{fig:translation-space-overhead-128gb}
     \end{subfigure}
     \begin{subfigure}{0.21\textwidth}
        \centering
        \includegraphics[scale=0.45]{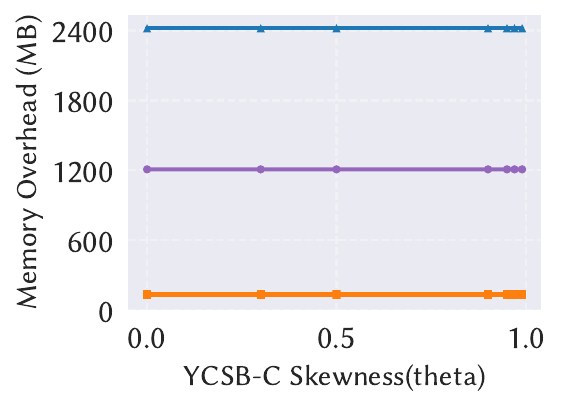}
        \subcaption{YCSB-C(614GB), 16GB Pool}
        \label{fig:translation-space-overhead-ycsbc}
     \end{subfigure}
     \hfill
    \begin{subfigure}{0.21\textwidth}
        \centering
        \includegraphics[scale=0.45]{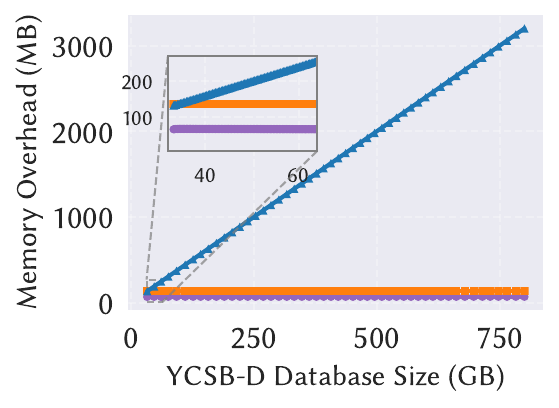}
        \subcaption{YCSB-D, 16GB Pool}
        \label{fig:translation-space-overhead-ycsbd}
     \end{subfigure}
    \Description{Four plots show translation memory overhead for Calico, hash tables, and vmcache across TPC-C, YCSB-C, and YCSB-D settings. Hash tables remain flat, vmcache grows with database size, and Calico stays closer to working-set-sized overhead when hole punching is effective.}
    \caption{
        \captionTitle{Translation Memory Overhead}
        \calico's hole-punching keeps translation space proportional to working set in TPC-C and
        YCSB-D. Hash tables maintain constant overhead, vmcache scales linearly with database size.
        YCSB-C shows challenging behavior with spatially dispersed hot keys preventing
        hole-punching, yet \calico outperforms vmcache by 2$\times$. YCSB-D demonstrates excellent
        hole-punching effectiveness (95.7\% reclamation) due to locality where older
        insertions become cold.
    }
    \label{fig:translation-space-overhead}
\end{figure}

\cref{fig:translation-space-overhead} shows translation memory as database size grows. Hash 
tables maintain constant overhead: 136 MB (16GB pool), 1,088 MB (128GB pool). vmcache grows linearly 
with database size. \calico's hole-punching mechanism reclaims memory for cold regions, exploiting 
temporal locality. For TPC-C with 16GB pool, \calico reclaims 86.2\% (1,379 MB of 1,600 MB), 
achieving 221 MB final usage (1.89$\times$ hash tables). At 128GB pool, reclamation drops to 
60.4\%, yielding 633 MB. YCSB-D (read-latest workload) demonstrates hole-punching effectiveness: as 
new records are inserted and become the most frequently accessed, older data transitions to cold 
state, enabling \calico to reclaim 95.7\% (1,532 MB of 1,600 MB), achieving just 68 MB final usage 
(0.50$\times$ hash tables, 46.9$\times$ less than vmcache). In contrast, YCSB-C presents a worst 
case: its read-only workload and Zipfian distribution spread hot keys across the entire keyspace. 
Even at high skewness (0.99), frequently accessed records scatter throughout the database, 
preventing hole-punching from creating contiguous cold regions. Across all skewness levels, \calico 
consumes 1,209 MB, yet still uses 2$\times$ less memory than vmcache.

\subsection{Ablation Study}

\begin{figure}[t]
    \centering
    \includegraphics[width=0.4\textwidth]{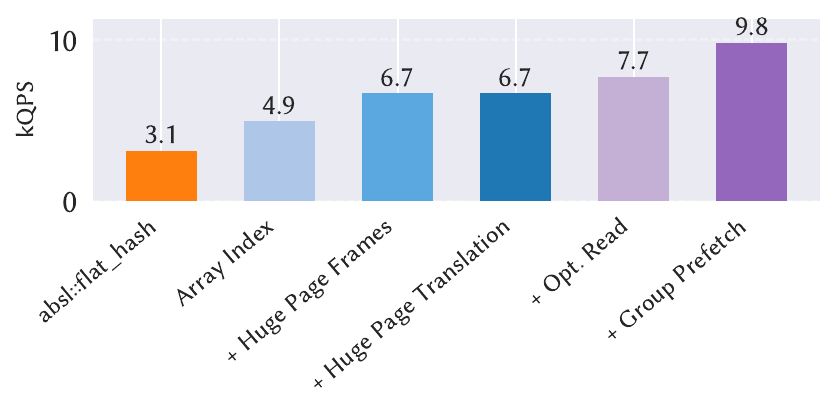}
    \Description{A bar chart presents the cumulative speedup of Calico optimizations on DEEP10M HNSW. Array indexing gives the largest gain, followed by huge pages, optimistic reads, and group prefetch.}
    \caption{
        \captionTitle{Ablation Study (HNSW DEEP10M, 0.88 Recall@10)}
        Cumulative speedup of \calico optimizations. Array indexing provides the 
        largest gain (1.59$\times$), followed by incremental benefits from huge pages, optimistic 
        reads, and prefetching, totaling 3.15$\times$ speedup over baseline.
    }
    \label{fig:ablation}
\end{figure}

We conduct an ablation study using the DEEP10M dataset and HNSW vector search to quantify the 
incremental benefits of \calico's optimizations when data fits in memory, shown in 
\cref{fig:ablation}. Array translation delivers the primary gain by 1.59$\times$, increasing 
throughput to 4.9~kQPS. Backing frame 
memory with 2MB huge pages adds a further 1.35$\times$ speedup by reducing TLB 
pressure, though applying huge pages to the compact translation array itself provides no additional 
benefit. Optimistic reads further improve performance to 7.7~kQPS by removing atomic reference counting overhead and enabling memory-level 
parallelism. Lastly, group prefetching reaches 9.8~kQPS by hiding memory latency during graph traversal, demonstrating that once translation 
overhead is minimized, memory latency becomes the dominant bottleneck that prefetching mitigates.

\begin{table}[t]
    \centering
    \caption{
        \captionTitle{Microarchitecture Analysis of Group Prefetch}
        In-memory graph BFS microbench on a 5M-node graph, comparing no prefetch vs.\ group
        prefetch for array translation, \texttt{absl::flat\_hash}, PrediCache, and vmcache
        (4KB/2MB). Group prefetch helps only array translation (1.201$\times$); hash-based and
        vmcache variants show little or negative gain. Per-node counters are reported as
        No$\rightarrow$Yes prefetch, and speedup is computed from average traversal time.
    }
    \label{tab:graph-prefetch-uarch}
    {\footnotesize \setlength{\tabcolsep}{3.2pt}
\begin{tabular}{lrrrr}
\toprule
\textbf{Mechanism} & \textbf{Speedup} & \textbf{Cycles/op} & \textbf{L3-stall/op} & \textbf{LLC-miss/op} \\
\midrule
\texttt{array} & 1.201$\times$ & 4903$\rightarrow$4082 & 5521$\rightarrow$4271 & 84.3$\rightarrow$65.8 \\
\texttt{absl::flat\_hash} & 1.008$\times$ & 11774$\rightarrow$11680 & 13736$\rightarrow$12835 & 158.7$\rightarrow$205.8 \\
\texttt{predicache} & 0.946$\times$ & 11201$\rightarrow$11840 & 11310$\rightarrow$11870 & 309.9$\rightarrow$173.9 \\
\texttt{vmcache(4KB)} & 0.916$\times$ & 5410$\rightarrow$5905 & 6295$\rightarrow$6650 & 93.7$\rightarrow$98.0 \\
\texttt{vmcache(2MB)} & 0.940$\times$ & 3657$\rightarrow$3878 & 3958$\rightarrow$4037 & 63.0$\rightarrow$71.9 \\
\bottomrule
\end{tabular}
}
\end{table}

\parhead{Why Group Prefetch Helps Array Translation More.} To isolate the impact of group prefetch, we run a focused in-memory graph BFS benchmark with 5M graph nodes. We compare no prefetch vs.\ group prefetch for
\calico's array translation, open-address hash translation, and PrediCache.
\cref{tab:graph-prefetch-uarch} shows that group prefetch helps only array translation
substantially (1.201$\times$), while open-address hash and PrediCache see no meaningful gain
(1.008$\times$ and 0.946$\times$). For \calico's array translation, group prefetch
reduces cycles/op by 16.7\%, L3-stall cycles/op by 22.6\%, and LLC misses/op by 22.0\%, which
translates to a clear end-to-end gain. In contrast, hash-based translation does not improve runtime
despite slightly lower cycles/op, and PrediCache regresses due to higher execution overhead. We
also include vmcache with 4KB and 2MB pages: both regress with group prefetch (0.916$\times$ and
0.940$\times$) and show higher cycles/op and L3-stall/op. The key structural difference is translation complexity on the prefetch path. In \calico, each PID
maps to one contiguous array entry, so high-fan-out graph traversal exposes many independent
translation loads that software prefetch can turn into useful memory-level parallelism and therefore hide such translation latency. In
hash-table and PrediCache designs, each prefetch operation still performs a hash-based lookup
(hash computation, probing/key checks, and possible collision handling) in order to prefetch its translation state.
These extra instructions and control dependencies reduce effective memory-level parallelism, so
prefetch overhead is not repaid by comparable latency hiding. vmcache similarly shows little benefit
because its state lookup is not the dominant dependent bottleneck on this workload. 
\begin{figure}[t]
     \begin{subfigure}{0.21\textwidth}
        \includegraphics[scale=0.5]{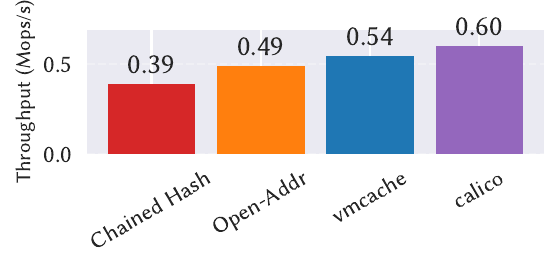}
        \subcaption{ARM Cortex-A72 (B-tree)}
        \label{fig:btree-arm}
     \end{subfigure}
     \hfill
     \begin{subfigure}{0.22\textwidth}
        \includegraphics[scale=0.5]{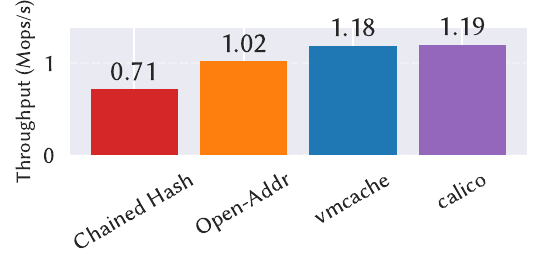}
        \subcaption{Intel Core i9-13900K (B-tree)}
        \label{fig:btree-intel}
     \end{subfigure}

     \begin{subfigure}{0.21\textwidth}
        \includegraphics[scale=0.5]{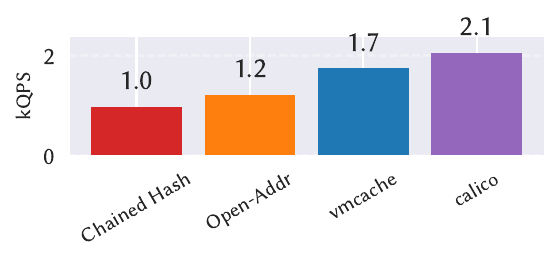}
        \subcaption{ARM Cortex-A72 (HNSW)}
        \label{fig:hnsw-arm}
     \end{subfigure}
     \hfill
     \begin{subfigure}{0.22\textwidth}
        \includegraphics[scale=0.5]{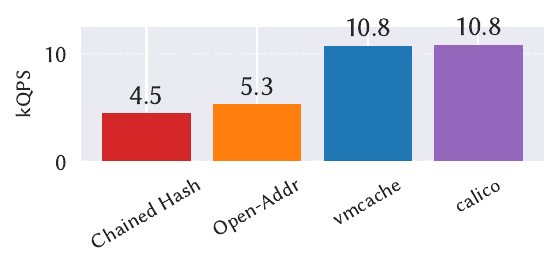}
        \subcaption{Intel Core i9-13900K (HNSW)}
        \label{fig:hnsw-intel}
     \end{subfigure}
    \Description{Four single-thread plots compare Calico, vmcache, and hash-table baselines on ARM and Intel for B-tree YCSB-C and HNSW. Calico matches or exceeds vmcache and remains above hash-table baselines across both architectures.}
    \caption{
        \captionTitle{Cross-Platform Validation using Single-threaded B+tree and HNSW vector search}
        \calico consistently outperforms hash tables and matches or exceeds vmcache across ARM and 
        Intel platforms.
    }
    \label{fig:cross-platform}
\end{figure}

\subsection{Cross-Platform Validation}
To validate that \calico's benefits generalize across CPU architectures, we conducted 
single-threaded experiments on ARM Cortex-A72 (16 cores, 32GB RAM) and Intel Core i9-13900K (24 
cores, 64GB RAM) using HNSW vector search with SIFT1M dataset and YCSB-C workloads with 100M 
128-byte records.

\parhead{HNSW Vector Search} (\cref{fig:hnsw-arm,fig:hnsw-intel}): On ARM, \calico achieves 
2~kQPS, outperforming vmcache by 18\% and chained hash by 2.1$\times$. On Intel, \calico reaches 
10.8~kQPS, matching vmcache and exceeding chained hash by 2.4$\times$. The larger ARM gap (18\% vs.\ 
1\%) suggests ARM is more sensitive to TLB overhead.

\parhead{B-tree YCSB-C} On ARM (\cref{fig:btree-arm,fig:btree-intel}), \calico achieves 
0.6~Mops/s, outperforming vmcache (4KB pages) by 10\% and open-addr hash by 22\%. On Intel, 
\calico reaches 1.19~Mops/s, exceeding open-addr hash by 17\%. \calico's huge pages (2MB) 
eliminate TLB pressure that vmcache faces with 4KB pages, while direct array indexing consistently 
outperforms hash tables (17--22\%) across both shallow (ARM) and deep (Intel) pipelines.

\section{RELATED WORK}

\parhead{Main-Memory DBMS and Buffer Management} Main-memory database 
systems~\cite{stonebraker07,kallman08,kemper11,diaconu13} manage data at tuple granularity and have 
been extended to larger-than-memory datasets~\cite{debrabant13,stoica13,Eldawy2014}. 
Anti-Caching~\cite{debrabant13} keeps indexes in memory while evicting cold tuples, limiting 
scalability~\cite{zhang2016reducing}. Recent buffer management research explored persistent 
memory~\cite{renen18,zhou2021spitfire}, remote 
memory~\cite{riekenbrauck2024three,hao2024towards,ziegler2022scalestore}, translation with SIMD hash 
table~\cite{liu2025scalecache}, or improving memory utilization under skew~\cite{zhou2023two, 
zhou2025tiered}. \calico complements these approaches: its direct array indexing delivers 
near-hardware translation performance at page granularity, enabling fine-grained eviction control 
without kernel modifications or eschewing buffer manager.


\parhead{Software Prefetching} Software prefetching has been widely studied to hide memory and 
IO latency in various contexts, including database query processing~\cite{chen2007improving, 
psaropoulos2017interleaving, jonathan2018exploiting} and transaction 
processing~\cite{he2020corobase, huang2023art}. Unlike existing work, \calico's group prefetch 
interface allows modern workloads such as graph-based vector search to express memory-level/IO 
parallelism to the buffer manager.

\parhead{Larger-Than-Memory Vector Search} Parallel SSD I/O has been exploited to improve 
disk-based vector search~\cite{shim2025turbocharging,wang2024starling,chen2021spann,jayaram2019diskann}. 
Unlike existing work, \calico focuses on improving in-memory performance of buffer-managed vector 
search index and I/O performance under memory pressure through group prefetch.

\parhead{OS-Level Memory Management} Recent research has explored improving memory management 
through OS-level mechanisms. Enhanced swapping systems~\cite{amaro2020can, jalalian2024extmem, 
gu2017efficient,wang2020semeru, zhong2023revisiting,papagiannis2021memory} aim to improve 
performance for larger-than-memory workloads, but operate at hardware page table granularity and 
face the eviction granularity problem. DBMS/OS co-design approaches such as 
libdbos~\cite{zhou2025practical} and CumulusDB~\cite{leis24vision} explore running database systems 
in privileged kernel space to enable more powerful abstractions for buffer management and 
snapshotting. In contrast, \calico demonstrates that careful user-space design can achieve 
hardware-page-table performance with fine-grained eviction without kernel modifications.

\section{CONCLUSION}

This paper presents \calico, a DBMS-managed array-translation buffer pool that closes a gap in the
buffer-management design space. Modern systems need one design that is simultaneously strong on
sequential scans, low-parallelism random lookups, high-parallelism random traversals, user-space
deployability, huge-page friendliness, and non-invasive integration with existing data structures.
Existing families satisfy only subsets of these dimensions. \calico combines direct array
translation, huge-page-backed frames, multi-level translation, hole punching, and group prefetch to
satisfy all six in one design.

Our evaluation shows that this design achieves hardware-competitive resident-page performance while
retaining DBMS control over eviction and I/O. \calico delivers 2--6$\times$ higher throughput than
page-table-based designs under memory pressure, 1.6$\times$ speedup over hash tables on B-tree
workloads, and 3.95$\times$ in-memory and 6.5$\times$ out-of-memory end-to-end speedup in
PostgreSQL/pgvector.


\bibliographystyle{ACM-Reference-Format}
\bibliography{ref}

\end{document}